\documentclass[twocolumn, prl, superscriptaddress,notitlepage]{revtex4-2}
\usepackage{graphicx}
\usepackage{dcolumn}
\usepackage{bm}
\usepackage[usenames,dvipsnames]{color}
\usepackage[most]{tcolorbox}
\usepackage{multirow}
\usepackage{gensymb}
\usepackage[normalem]{ulem}
\usepackage{CJK}
\usepackage{comment}
\usepackage[colorlinks, linkcolor=blue,anchorcolor=blue,citecolor=blue,urlcolor=blue]{hyperref}
\usepackage{amssymb}
\usepackage{pifont}
\usepackage{physics}
\usepackage{natbib}
\usepackage{xcolor}

\usepackage{color,soul}

\begin{document}
\begin{CJK*}{UTF8}{}

\title{Spin squeezing by geometric focusing in vacuum Rabi oscillations}
\author{Aoqin Liang}
\affiliation{International Center for Quantum Materials, School of Physics, Peking University, Beijing 100871, China}
\affiliation{Beijing Key Laboratory of Quantum Devices, Peking University, Beijing 100871, China}

\author{Qi Liu}\email[]{qiliu23@mit.edu}
\affiliation{National Laboratory of Solid State Microstructures and School of Physics, Collaborative Innovation Center of Advanced Microstructures, Nanjing University, Nanjing 210093, China}
\affiliation{
   MIT-Harvard Center for Ultracold Atoms and Research Laboratory of Electronics, Massachusetts Institute of Technology, Cambridge, Massachusetts 02139, USA}

\author{Guoqing Wang} \email[]{gq\_wang@pku.edu.cn}
\affiliation{International Center for Quantum Materials, School of Physics, Peking University, Beijing 100871, China}
\affiliation{Beijing Key Laboratory of Quantum Devices, Peking University, Beijing 100871, China}

\begin{abstract}
We show that vacuum Rabi oscillations can directly generate spin squeezing through geometric focusing on the Bloch sphere. Starting from a coherent spin state resonantly coupled to a cavity initially in the vacuum state, quantum fluctuations are focused by the curvature of the Bloch sphere as the collective spin approaches the atomic ground state, producing squeezing transverse to the direction of motion. The squeezing timescale is set by the collective Rabi frequency $t_s\sim 1/(g\sqrt{N})$. The optimal Wineland squeezing parameter scales as $\xi_{\rm opt}^2\propto N^{-1/3}$, which is an outcome of the competition between the geometric focusing effects and the cavity-field vacuum fluctuations. The squeezing remains robust against realistic dissipation.  In the end, an application example of $^{171}$Yb is briefly discussed to show the feasibility of our protocol.
\end{abstract}

\maketitle

\end{CJK*}

\textit{Introduction.---} Quantum sensing and metrology have led to broad applications ranging from timekeeping, magnetometry, inertial sensing, gravitational wave detection, and dark matter search~\cite{ludlow_sr_2008,zaporski_quantumamplified_2025,gustavson_precision_1997,durfee_longterm_2006,mohr_codata_2008,pelluet_atom_2025,pedrozo-penafiel_entanglement_2020,herb_quantum_2025}. 
While the sensitivity can be improved by scaling up the system size with a standard quantum limit of $1/\sqrt{N}$, quantum correlations such as entanglement and squeezing can further boost the sensitivity, ideally towards the ultimate Heisenberg limit $1/N$~\cite{giovannetti_quantum_2006,wineland_spin_1992,wineland_squeezed_1994,ma_quantum_2011,louchet-chauvet_entanglementassisted_2010,hosten_measurement_2016}. 
Spin squeezing can be generated by nonlinear spin-spin interactions such as the one-axis-twisting (OAT) and two-axis-twisting (TAT) Hamiltonians~\cite{kitagawa_squeezed_1993}, which can either be engineered by native interactions such as atomic collisions ~\cite{hamley_spinnematic_2012,bohnet_quantum_2016,lucke_twin_2011,riedel_atomchipbased_2010} or dipolar interactions~\cite{wu_spin_2025}, or be mediated by light-matter interactions in cavity-coupled systems~\cite{schleier-smith_squeezing_2010,leroux_implementation_2010,braverman_nearunitary_2019}.

Typical cavity-based spin squeezing employs a cavity probe, either in the cavity-feedback regime where dispersive atom-photon coupling generates an effective OAT interaction~\cite{Li2022,leroux_implementation_2010,braverman_nearunitary_2019,schleier-smith_squeezing_2010}, or in the quantum nondemolition (QND) regime where the detection of a transmitted photon projects the system onto a squeezed state~\cite{chen_conditional_2011,schleier-smith_states_2010,ma_quantum_2011,Chen2015}. It was later shown that an electromagnetic vacuum can serve as a resource for generating squeezed states, eliminating the need for a cavity probe~\cite{hu_vacuum_2017}. More recently, a dissipative approach utilizing dark states in multilevel atoms was proposed, where spin squeezing is generated via the process of collective decay~\cite{sundar_squeezing_2024}. Conventional coherent cavity-squeezing protocols based on effective spin-spin interactions typically operate in a dispersive, weak-excitation regime, where the cavity field can be adiabatically eliminated and second-order spin-spin interactions dominate the dynamics. Recent work has demonstrated that resonant spin-boson coupling can generate fast squeezing without relying on such second-order interactions~\cite{barberena_fast_2024}.

In this Letter, we reveal a geometric mechanism for spin squeezing arising directly from vacuum Rabi oscillations, during which a two-level atomic ensemble resonantly exchanges energy with a cavity field according to the Tavis-Cummings model~\cite{rose_coherent_2017,tavis_exact_1968}. Starting from a coherent spin state (CSS), the collective spin follows pendulum-like dynamics~\cite{kaluzny_observation_1983}, while the curvature of the Bloch sphere focuses quantum fluctuations as it approaches the atomic ground state, producing squeezing orthogonal to the direction of motion. The competition between this geometric focusing and the cavity-field vacuum fluctuations determines an optimal initial angle and yields the scaling of the Wineland squeezing parameter $\xi_{\rm opt}^2\propto N^{-1/3}$. The squeezing develops on the collective Rabi timescale $t_s\sim 1/(g\sqrt{N})$, set by the first-order light-matter coupling rather than an effective second-order spin-spin interaction. We characterize its robustness against dissipation and find that pronounced squeezing persists for experimentally accessible cavity parameters. For $10^5$ cavity-coupled $^{171}$Yb atoms, we predict 15 dB of squeezing within 1~ns at a cavity cooperativity $\eta=100$. 

\begin{figure*}[t]
    \centering
    \includegraphics[width=\textwidth]{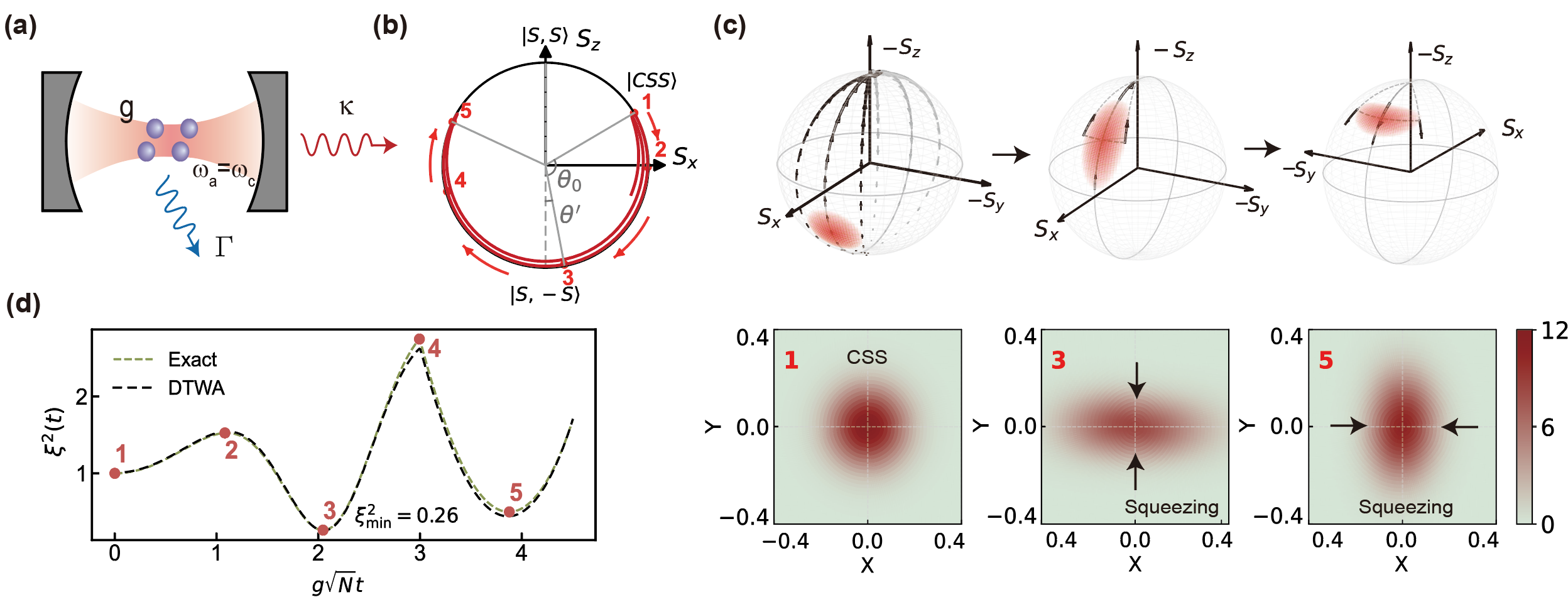}
    \caption{(a) Atom-cavity coupled system. The single-atom vacuum Rabi frequency is $2g$, while the linewidths of the atomic transition and cavity are $\Gamma$ and $\kappa$, respectively. Our protocol operates under the atom-cavity resonance scenario ($\omega_{c}=\omega_{a}$). (b) The initial state is prepared in a CSS lying in the $z$-$x$ plane, with an angle $\theta_0$ relative to $-z$ axis. The red curve shows the trajectory of the mean spin vector during vacuum Rabi oscillation. Several characteristic points are highlighted by red dots, labeled 1-5. (c) The Husimi $Q$ functions at point 1, 3, and 5. Arrows on the surface of the Bloch sphere indicate the local velocity vectors. The meridians of the Bloch sphere converge at the $-z$ pole ($S_z=-S$), leading to spin squeezing at point 3. A second squeezing feature appears at point 5 due to velocity variations between trajectories at different latitudes. The coordinates are defined as $X=\sin(\theta)\cos(\phi)$ and $Y=\sin(\theta)\sin(\phi)$, where $\theta$ and $\phi$ are the polar and azimuthal angle of the Bloch vector, respectively. (d) Wineland squeezing parameter $\xi^2$ as a function of evolution time $g\sqrt{N}t$, calculated using both exact diagonalization and the DTWA method, with $N=150$ atoms and an initial CSS polar angle of $\theta_0=120.6^{\circ}$.}
    \label{fig:1}
\end{figure*}
\textit{Principle.---} We consider a general cavity-coupled many-atom system as shown in Fig.~\ref{fig:1}(a), which can be described by the Tavis-Cummings (TC) model
\begin{equation}
\hat{H}_{\rm TC}=\omega_ca^{\dagger}a+\omega_a S_z+ g(aS_++a^{\dagger}S_-),
\label{Eq: T-C model}
\end{equation}
where $S_i=\sum_{k=1}^{N}\sigma_k^i/2~(i=x,y,z)$ denotes the collective spin operator with $\sigma^i_k$ the Pauli operator of the $k$-th particle. $a^\dagger (a)$ stands for the creation (annihilation) operator of the cavity field, while $\omega_c~(\omega_a)$ represents the cavity (atomic transition) resonance frequency. Here, we consider the strong coupling regime $g\sqrt{N}\gg\Gamma+\kappa$ under the resonance condition $\omega_a=\omega_c$. The initial state is prepared in a CSS with a vacuum cavity field (point 1 in Fig.~\ref{fig:1}(b))
\begin{equation}
\ket{\rm CSS}\otimes|{\rm vac}\rangle=\left(\cos\frac{\theta_0}{2}|\downarrow\rangle+\sin\frac{\theta_0}{2}|\uparrow\rangle\right)^{\otimes N}\otimes|\rm vac\rangle.
\label{Eq:CSS}
\end{equation}
To start with, we neglect dissipation in the system and evaluate the fully coherent dynamics of vacuum Rabi oscillation, which conserves the total excitation number $a^{\dagger}a+S_z$. The evolution of the polar angle $\theta(t)$ is described by $\ddot{\theta}(t)+g^2N\sin\theta(t)=0$~\cite{kaluzny_observation_1983}, which is identical to an oscillating pendulum as shown in Fig.~\ref{fig:1}(b). The Bloch vector passes through points 1-5 in the first period. In the following, we show that the first squeezing position is located at point 3 near the atomic ground state, while the second squeezing point is located at point 5 near the other highly excited atomic state.

The squeezing near the atomic ground state can be understood as a geometric focusing effect arising from the curvature of the Bloch sphere. Initially in the CSS state, the quasi-probability distribution of the atomic state exhibits an isotropic Gaussian shape as shown in Fig.~\ref{fig:1}(c). Each sampling point of the distribution moves along its corresponding meridian on the Bloch sphere surface. When approaching the atomic ground state ($-z$ pole), neighboring meridians converge, focusing the trajectory bundle and suppressing fluctuations along the transverse direction. This geometric focusing produces the transverse squeezing along the $y$ direction at point 3; after passing the $-z$ pole, the same curvature causes transverse defocusing and antisqueezing. Along the meridian direction, the squeezing and antisqueezing dynamics are governed by the superradiance rate, which depends on the polar angle. The initial photon emission rate is described by $\ddot{n}(0)\propto g^2N^2\sin^2(\theta_0)$, hence, the sampling points that are closer to the $+z$ pole (with larger $\theta_0\sim180^\circ$) evolve with a slower speed. This rate difference stretches (contracts) the quasi-probability distribution, leading to anti-squeezing (squeezing) along the meridian at point 3 (5) as shown in Fig.~\ref{fig:1}(c).

\begin{figure}[t]
    \centering
    \includegraphics[width=0.495\textwidth]{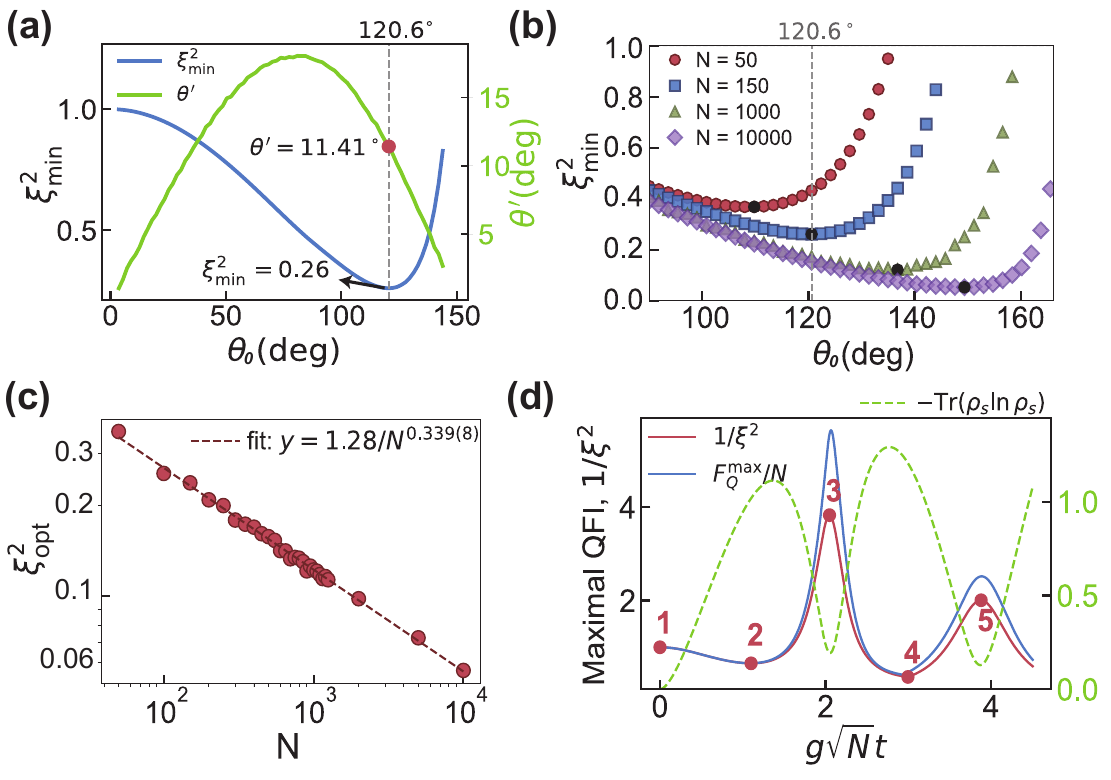}
    \caption{(a) Minimum squeezing parameter $\xi^2_{\text{min}}$ and the optimal Bloch vector position $\theta^\prime$ (right $y$-axis) as a function of initial angle $\theta_0$. $N=150$ is used in the calculations based on DTWA.
    (b) $\theta_0$-dependence of the minimum squeezing parameter for different atom number $N$. As $N\rightarrow\infty$, optimal $\theta_0$ moves towards $180^{\circ}$ while $\theta'$ approaches $0$.
    (c) Scaling of the squeezing parameter optimized over $\theta_0$ and $\theta'$ with atom number $N$. (d) Evolution of the maximal QFI, inverse of the squeezing parameter and entanglement entropy (right $y$ axis) for $N=150$ and $\theta_0=120.6^{\circ}$. Spin-photon entanglement nearly vanishes at the locally optimal squeezing points 3 and 5.}
    \label{fig:2}
\end{figure}
\textit{Squeezing dynamics and scaling.---}
To analyze the dynamical process of squeezing generation, we employ the discrete truncated Wigner approximation (DTWA), which has proven powerful for analyzing driven-dissipative many-body dynamics with high precision ~\cite{hosseinabadi_userfriendly_2025,polkovnikov_phase_2010,schachenmayer_manybody_2015a}. The core idea of the DTWA is to map the high-dimensional quantum state $\rho$ onto a discrete Wigner distribution in phase space constructed from the classical degrees of freedom. 
The complete quantum dynamics is efficiently computed via an ensemble distribution of a set of classical evolution trajectories, accounting for both the quantum uncertainty in the initial state as well as the effects of damping and noise during the evolution~\cite{hosseinabadi_userfriendly_2025}.

In the absence of dissipation, the dynamics of TC model is confined by permutation symmetry to the $(N+1)$-dimensional symmetric Dicke subspace, enabling an exact numerical solution owing to the conservation of total excitations~\cite{tavis_exact_1968}. 
The initial CSS state in the phase space shows a Gaussian distribution $W(\mathbf{s},\alpha,\mathbf{\mu}, N)
= \frac{2}{\pi N}\exp\left(-\frac{2}{N}s_{\perp}^2\right) \delta\left(\mathbf{\mu}\cdot(\mathbf{s}-\mathbf{\mu})\right) \times \frac{2}{\pi}\exp\left(-2|\alpha|^2\right)$, where $\mathbf{\mu}=N/2\left(\sin\theta_0,0,-\cos\theta_0\right)$ is the initial mean spin vector and $s_{\perp}$ is the transverse spin component with an isotropic variance of ${N}/{4}$. The intracavity field has a mean amplitude of $\alpha=0$ and exhibits vacuum fluctuations with a variance of $1/4$.
The evolution of a single spin trajectory under the TC Hamiltonian is then described by $d\alpha^*/dt=igs_+$, $ids_+/dt=2gs_z\alpha^*$, and $ds_z/dt=2g\Im(\alpha S_+)$, with $s_+=s_x+is_y$ denoting the spin coherence, consistent with the Maxwell-Bloch equations describing the mean-field evolution of the Bloch vector. 
Considering that the initial intracavity field is in the vacuum state, the emitted photon is phase-locked to the spin, causing the single spin trajectory to follow the meridian through its initial position and behave like a pendulum.

In Fig.~\ref{fig:1}(d), we show the evolution of the Wineland squeezing parameter $\xi^2=N\left(\Delta\mathbf{S}_{\mathbf{n}}\right)^2/{|\mathbf{S}|^2}$~\cite{wineland_spin_1992,wineland_squeezed_1994,sorensen_entanglement_2001}, with $\mathbf{n}$ denoting the direction of minimal spin variance. The points 1-5 correspond to the locations labeled in Fig.~\ref{fig:1}(b). When the Bloch vector passes point 2 near the equator, the spin noise in both $X$ and $Y$ directions is anti-squeezed. Here, we define the coordinates $X=\sin(\theta)\cos(\phi)$ and $Y=\sin(\theta)\sin(\phi)$, with $\theta$ and $\phi$ denoting the polar and azimuthal angle of the Bloch vector, respectively. As the vector moves closer to the $-z$ pole (point 3), the optimal squeezing is obtained along the $Y$ quadrature with $\xi^2_{\rm min}=0.26$, resulting from the geometric focusing. After passing the pole, the velocity difference along the meridian reverses sign, and the curvature of the Bloch sphere drives the bundle of trajectories apart, leading to increased spin fluctuations along both the $X$ and $Y$ quadratures. The direction of minimal variance switches from $Y$ to $X$ at point 4, and a second local mininum of the squeezing parameter is reached at point 5 along the $X$ direction.

We evaluate the minimal squeezing parameter $\xi_{\rm min}^2$ as a function of the initial angle $\theta_0$ in Fig.~\ref{fig:2}(a). One can see that the optimal angle lies between $90^\circ$ and $180^\circ$, which shifts towards $180^\circ$ as $N\rightarrow\infty$ (Fig.~\ref{fig:2}(b)). The nonmonotonic dependence on $\theta_0$ originates from a competition between geometric focusing and vacuum fluctuations of the intracavity field. In the absence of vacuum fluctuations, the isotropic spin distribution would evolve into an $\infty$ shape at the $-z$ pole, where the spin variance along the $Y$ axis is approximately given by the squared product of antisqueezed meridian length $\Delta l\sim \sqrt{N}[\cos^{-1}\frac{\theta_0}{2}E(\sin\frac{\theta_0}{2})-\cos\frac{\theta_0}{2}K(\sin\frac{\theta_0}{2})]$ and the spanned longitude angle $\Delta \phi\sim1/(\sqrt{N}\sin\theta_0)$, as derived in the Supplemental Material (SM)~\cite{SOM}. Here, $K$ and $E$ denote the complete elliptic integrals of the first and second kinds, respectively. The Wineland squeezing parameter can be approximated by $(\Delta\phi\Delta l)^2/N$, which increases monotonically with $\theta_0$ and decreases with $N$~\cite{SOM}. As $\theta_0$ approaches $\pi$, the squeezing parameter scales as $4/(N\delta^4)$, with $\delta=\pi-\theta_0$.

Although the geometric focusing effect prefers smaller $\theta_0$ for better spin squeezing, vacuum fluctuation of the intracavity field plays a competing role. At the beginning of the evolution, vacuum fluctuation drives the spin along a random transverse axis, leading to deviations from its initial meridian. As the spin evolves down towards the $-z$ pole, the photon emitted into the cavity mode tends to lock the spin trajectory to a single meridian. The total light field emitted by the spin ensemble is upper-bounded by $\sqrt{N}\sin(\theta_0/2)$~\cite{SOM}, thus increasing $\theta_0$ towards $\pi$ improves the locking performance, enabling stronger geometric focusing as the system evolves towards the $-z$ pole.  In Fig.~\ref{fig:2}(b), we find that for a fixed initial angle $\theta_0=120.6^\circ$ (gray dashed line), the squeezing parameter improves with increasing $N$ but eventually saturates at a finite value, limited by the intracavity vacuum fluctuations. To achieve scalable spin squeezing, one needs to increase $\theta_0$ with $N$ to mitigate the contribution from the vacuum fluctuation. With the cumulant-expansion treatment (see a detailed analysis in the SM~\cite{SOM}), we find that the squeezing parameter contributed by the vacuum fluctuation is approximately $\delta^2/8$, in contrast to $4/(N\delta^4)$ arising from geometric focusing. Minimizing the total contribution yields $\delta^2/8+4/(N\delta^4)\geq(3/4)N^{-1/3}$ and an optimal initial angle $\theta_0=\pi-\delta_{\text{opt}}=\pi-2N^{-1/6}$, revealing the $N^{-1/3}$ scaling as a direct consequence of the competition between geometric focusing and cavity vacuum fluctuations. Numerical simulation in Fig.~\ref{fig:2}(c) shows that the optimal squeezing parameter follows a scaling law of $N^{-0.339(8)}$, consistent with our analysis.

In Fig.~\ref{fig:2}(d), we further compare the evolution of quantum Fisher information (QFI) $\max_{\vec{n}}{F_Q[\hat{\rho}, S_{\vec{n}}]}$ for $N=150, \theta_0=120.6^{\circ}$ with the inverse squeezing parameter $1/\xi^2$. We find qualitatively similar dynamics with both quantities exhibiting local optima at points 3 and 5, clearly surpassing the value of the CSS state (point 1). The nearly identical values at the optimal squeezing points also indicate a near-unitary squeezing generation. This is further supported by the dynamics of entanglement entropy, which reaches local minima close to 0 at points 3 and 5 (right $y$ axis), indicating weak entanglement between the atoms and intracavity photons.

Remarkably, the vacuum Rabi oscillation also generates squeezing of the cavity field, revealing a simultaneous reshaping of quantum fluctuations in the spin and light sectors. At the two successive spin squeezing events, the cavity field noise is squeezed along two orthogonal field quadratures over the course of the coherent emission and reabsorption of cavity photons. The optimal light-mode squeezing occurs simultaneously with the spin squeezing when the spin-light correlations approach zero. A detailed discussion is provided in the SM~\cite{SOM}.

\begin{figure}[t]
    \centering
    \includegraphics[width=0.48\textwidth]{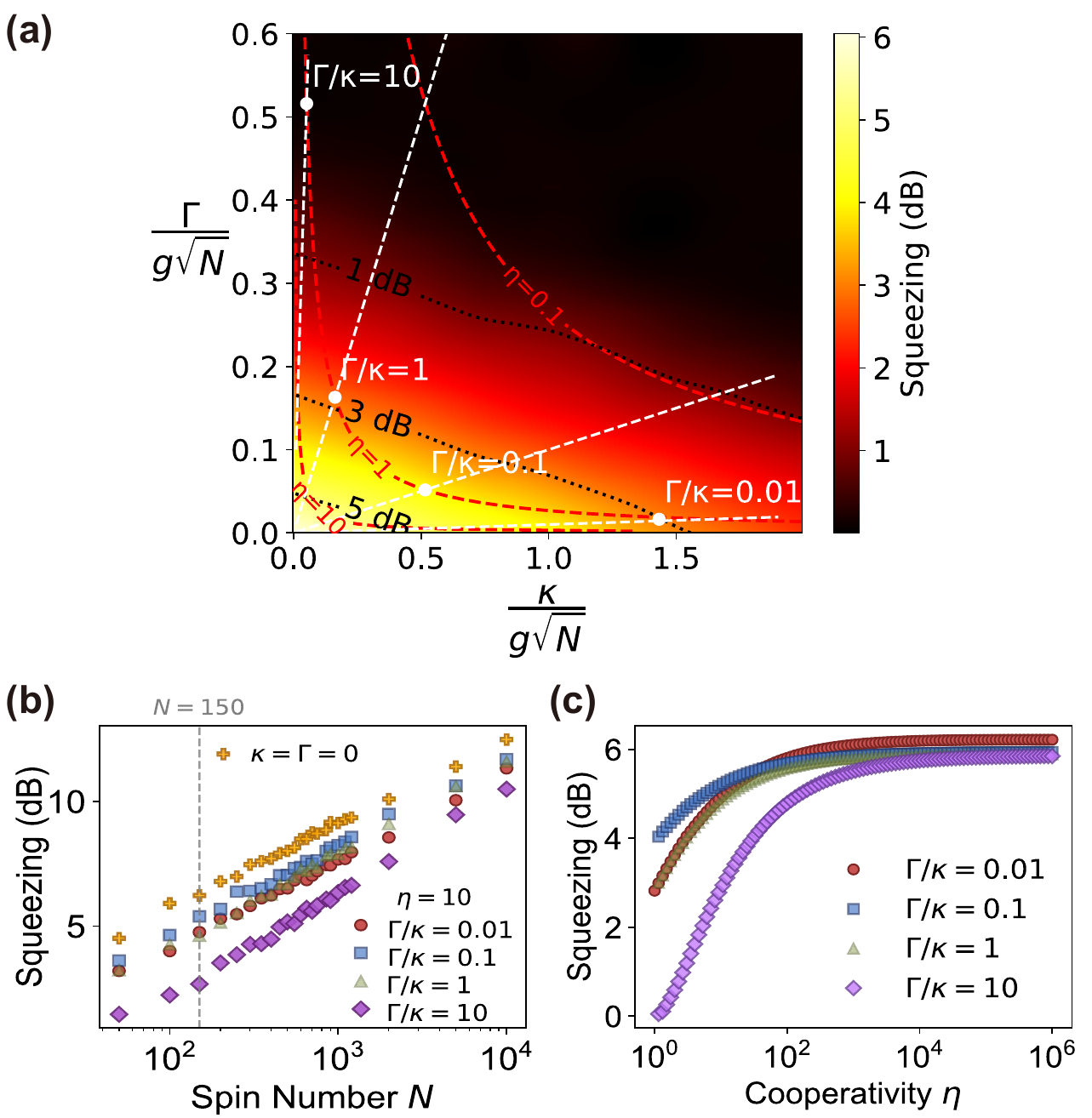}
    \caption{Effects of dissipation. (a) Optimal squeezing parameter (at point 3) as a function of atomic and cavity linewidths with $N=150$ and $\theta_0=120.6^{\circ}$. 
    The contours are shown in black dashed curves and the equal-cooperativity lines are illustrated in red dashed curves. (b) Scaling of spin squeezing with atom number $N$ for different dissipation ratios with fixed cooperativity of $\eta=10$. The ideal performance without dissipation is denoted by the yellow points as a comparison. For each $N$, the optimal initial angle $\theta_0$ is obtained from the calculations without considering dissipation. (c) Squeezing parameter as a function of $\eta$ for different dissipation ratios with $N=150$ and $\theta_0=120.6^{\circ}$.
    }
    \label{fig:3}
\end{figure}

\textit{Effects of dissipation.---}Now we study the dissipation effects including the atomic spontaneous emission $L_{i\Gamma}=\sqrt{\Gamma}\ket{\downarrow_i}\bra{\uparrow_i}$ and cavity loss $L_a=\sqrt{\kappa}a$. 
In Fig.~\ref{fig:3}(a), we show the squeezing parameter as a function of both $\Gamma$ and $\kappa$ for $N=150, \theta_0=120.6^{\circ}$. Notably, the squeezing persists even when the strong coupling condition $g\sqrt{N}\gg\Gamma+\kappa$ is not strictly satisfied, particularly in the case of strong cavity loss $\kappa\gtrsim g\sqrt{N}$. One can see that the squeezing parameter is more sensitive to atomic spontaneous emission than to cavity loss, indicating that spontaneous emission is more detrimental than cavity photon leakage. We also draw a few equal-cooperativity lines for $\eta=4g^2/\kappa\Gamma=0.1,1,10$. In general, larger cooperativity favors stronger squeezing, while the ratio of $\Gamma/\kappa$ also plays an important role. In Fig.~\ref{fig:3}(b), we calculate the scaling law under various dissipation conditions, which shows that the squeezing remains scalable under dissipation. It is also worth noting that the squeezing benefits from a ratio $\Gamma/\kappa$ slightly smaller than 1. Figure~\ref{fig:3}(c) further shows that, for $N=150, \theta_0=120.6^{\circ}$, an optimal dissipation ratio of $\Gamma/\kappa=0.1$ results in nearly 6~dB squeezing at a moderate cooperativity of $\eta=10$, whereas achieving comparable squeezing requires a much higher cooperativity of $\eta\sim 1000$ for a larger ratio $\Gamma/\kappa=10$. Our analysis thus provides a qualitative approach for designing cavity parameters for practical applications.

\begin{figure}[t]
    \centering
    \includegraphics[width=0.48\textwidth]{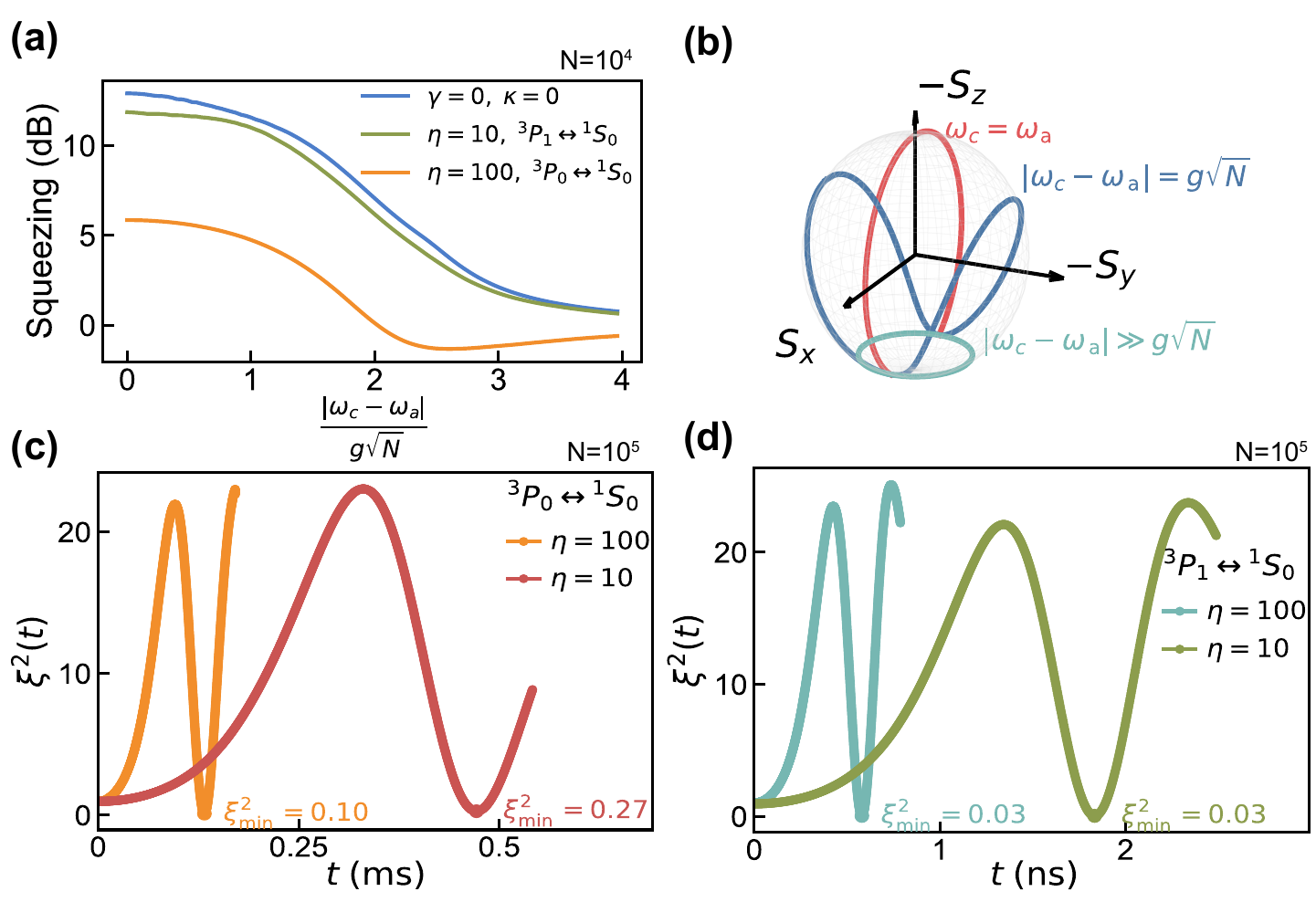}
    \caption{(a) Squeezing parameter as a function of atom-cavity detuning. We consider two transitions in the $^{171}$Yb atoms, ${^1}S_0\leftrightarrow {^3}P_0$ and ${^1}S_0\leftrightarrow{^3}P_1$, with selected cavity linewidths $\kappa/2\pi=1$~kHz and $\kappa/2\pi=1.82$~MHz, respectively. Parameters used in the calculation include atom number of $N=10^4$, initial angle of $\theta_0=149.4^{\circ}$, and cavity cooperativity $\eta=10,100$. (b) Evolution trajectory for the on-resonance (red curve), critical detuning ($|\omega_c-\omega_a|=g\sqrt{N}$, blue curve) and large detuning ($|\omega_c-\omega_a|\gg g\sqrt{N}$, cyan curve) cases. (c-d) Evolution of $\xi^2$ on the two atomic transitions. We use atom number $N=10^5$, initial angle $\theta_0=163.2^{\circ}$, and vanished atom-cavity detuning.}
    \label{fig:4}
\end{figure}

\textit{Application.---} Here, we discuss an exemplary application to the ${}^ {171}\mathrm{Yb}$ optical lattice clock. There are two potential choices of cavity-coupled energy levels: $^{1}S_0$ and $^{3}P_0$ coupled by the narrow linewidth clock transition ($\Gamma/2\pi=7.6~\rm mHz$), or $^{1}S_0$ and $^{3}P_1$ coupled by the intercombination transition ($\Gamma/2\pi=182~\rm kHz$). Accordingly, we choose cavity linewidths of $\kappa/2\pi=1$~kHz and $\kappa/2\pi=1.82$~MHz for the two schemes, respectively. The operational procedure on the clock transition is described as follows. The atomic ensemble is initialized in a CSS state and undergoes vacuum Rabi oscillation, during which the optimal squeezed state is obtained when the system approaches the ground state. We then rotate the squeezed state back to the equator with a $\pi/2-\theta^\prime$ pulse around the $y$-axis, followed by a free evolution duration $\tau_R$ and a $\pi/2$ pulse to implement the Ramsey spectroscopy. The measurement of spin populations is then performed with a reduced quantum noise set by the squeezing parameter.

For the narrow clock transition, a major concern of our squeezing scheme is the robustness against atom-cavity detuning. In Fig.~\ref{fig:4}(a), we calculate the achieved squeezing as a function of atom-cavity detuning with $N=10^4, \theta_0=149.4^{\circ}$ and $\eta=10,100$.  Our calculation shows that although the squeezing decreases with increasing detuning, it remains nearly flat for detuning much smaller than the collective coupling strength $g\sqrt{N}$. In the case of nonzero detuning, the spin trajectory becomes a saddle line from the pendulum circle, as shown in Fig.~\ref{fig:4}(b). We note that in the large detuning limit, the evolution reduces to a single circle wrapping around the $z$ axis with weak photon exchange with the vacuum. In this scenario, one can adiabatically eliminate the light field to obtain the effective OAT Hamiltonian to generate spin squeezing, albeit at a much slower rate~\cite{hu_vacuum_2017}. 
The evolution of the squeezing parameter with an atom number $N=10^5$ at initial angle $\theta_0=163.2^{\circ}$ is shown in Figs.~\ref{fig:4}(c-d). With $\eta=100$, we can reach 10~dB squeezing on the clock transition within 0.2~ms and 15~dB on the broader intercombination transition within 1~ns. With limited cavity cooperativity, one can prepare a squeezed state on the broad transition first with a fast squeezing rate, and then transfer the squeezing to the clock transition for optical clock application.

\textit{Conclusion.---} In conclusion, we have identified a geometric mechanism for spin squeezing arising directly from vacuum Rabi dynamics. As a coherent spin state approaches the atomic ground state, the curvature of the Bloch sphere focuses neighboring spin trajectories and suppresses quantum fluctuations transverse to the direction of motion.
Cavity vacuum fluctuations compete with this geometric focusing, selecting an optimal initial angle and giving rise to the scaling $\xi_{\text{opt}}^2\propto N^{-1/3}$. The accompanying light-mode squeezing further reveals that vacuum Rabi dynamics coherently reshapes quantum fluctuations in both matter and light. The spin squeezing remains robust against dissipation and atom-cavity detuning over experimentally accessible parameter regimes.  Crucially, squeezing develops on the collective Rabi timescale $1/(g\sqrt{N})$, set directly by the first-order light-matter coupling rather than an effective second-order interaction.   For a cavity-coupled ensemble of $10^5$ $^{171}$Yb atoms with cooperativity $\eta=100$, we predict that 15~dB of squeezing can be prepared within $1$~ns on the ${}^{3}P_1\leftrightarrow{}^{1}S_0$ transition.
Our results establish geometric focusing during vacuum Rabi oscillations as a distinct mechanism for generating spin squeezing.

\acknowledgements
We acknowledge Vladan Vuleti\'c, Yongchun Liu, Zeyang Li, Haoyang Gao , Hongzheng Zhao, Shixuan Zhang, Xin Tian for fruitful discussions. This work was supported by the Fundamental Research Funds for the Central Universities, Peking University.

\bibliography{main} 

\onecolumngrid
\clearpage
\begin{center}
\textbf{\large Supplemental Material: Spin squeezing by geometric focusing in vacuum Rabi oscillations}
\end{center}
\setcounter{section}{0}
\setcounter{equation}{0}
\setcounter{figure}{0}
\setcounter{table}{0}
\setcounter{page}{1}
\makeatletter
\renewcommand{\theequation}{S\arabic{equation}}
\renewcommand{\thesection}{S\arabic{section}}
\renewcommand{\thefigure}{S\arabic{figure}}


\tableofcontents

\tableofcontents

\section{Semiclassical dynamics and squeezing}

\subsection{Model}

Throughout this work, we discuss the Tavis--Cummings (TC) model describing a cavity-coupled ensemble of two-level atoms,
\begin{equation}
H_{TC}=\omega_ca^\dagger a+\omega_aS_z+g(aS_+ +a^\dagger S_-),
\end{equation}
where $\omega_c$ is the frequency of the cavity mode, $2g$ is the single-atom vacuum Rabi frequency, and $S_z$ and $S_\pm=S_x\pm iS_y$ are collective spin operators.

In the absence of dissipation, permutation symmetry confines the TC model to the $(N+1)$-dimensional symmetric Dicke subspace. Because the total excitation number is conserved, an exact numerical solution is feasible. Nevertheless, we continue our presentation within the DTWA framework\cite{hosseinabadi_userfriendly_2025}, which provides an intuitive understanding of the evolution. In this framework, the TC model is treated in its classical form as $H_{TC}=\omega_c({\rm Re}[\alpha]^2+{\rm Im}[\alpha]^2 )+\omega_aS_z+2g({\rm Re}[\alpha]S_x-{\rm Im}[\alpha]S_y )$, with the spin and light-mode operators directly replaced by their corresponding complex numbers.

The initial CSS in the phase space spanned by ($S_x$, $S_y$, $S_z$, $\Re(\alpha)$, $\Im(\alpha)$) has a Gaussian distribution,
\begin{equation}
W(\vec{S},\alpha,\vec{\mu},N)
= \frac{2}{\pi N}\exp\left(-\frac{2}{N}S_{\perp}^2\right)
   \delta\left(\hat{\mu}\cdot(\vec{S}-\vec{\mu})\right) \times \frac{2}{\pi}\exp\left(-2|\alpha|^2\right),
\label{eq:4}
\end{equation}
where $\vec{\mu}=N/2\left(\sin(\theta_0),0,-\cos(\theta_0)\right)$ is the initial mean-spin vector and $S_{\perp}$ is the projection of $\vec{S}$ onto the plane perpendicular to $\vec{\mu}$. The Wigner function indicates that the initial spin quasiprobability is distributed isotropically in the plane perpendicular to the mean-spin vector, with variance ${N}/{4}$. It also gives the expectation value $\langle \hat{S}^2\rangle={S(S+1)}$ for $S={N}/{2}$. The light mode has vacuum fluctuations around the mean value $\alpha=0$, with variance $1/4$ along each quadrature.

The DTWA method shows that the evolution of the Wigner function under the TC Hamiltonian behaves like an incompressible fluid in phase space and follows the classical Liouville equation,
\begin{equation}
    \frac{\partial W}{\partial t}=\{H_{TC},W\}_p+O(corrections),
    \label{eq:liouville}
\end{equation}
where $\{,\}_p$ is the classical Poisson bracket defined as $\{O,G\}_p=\left(\vec{S}\times\frac{\partial O}{\partial \vec{S}}\right)\cdot\frac{\partial G}{\partial\vec{S}}+\frac{1}{2}(\frac{\partial O}{\partial Im[\alpha]}\frac{\partial G}{\partial Re[\alpha]}-\frac{\partial O}{\partial Re[\alpha]}\frac{\partial G}{\partial Im[\alpha]})$. The correction terms are negligible in the large-$N$ limit. Under the resonant condition, we enter a rotating frame in which both $\omega_c$ and $\omega_a$ disappear. The classical canonical equations for a single trajectory are then
\begin{equation}
\begin{split}
&\frac{d\alpha^*}{dt}=igS_+\\
&i\frac{dS_+}{dt}=2gS_z\alpha^*\\
&\frac{dS_z}{dt}=2g\Im[\alpha S_+],
\label{eq:classical}
\end{split}
\end{equation}
where $S_+=S_x+iS_y$ is the spin coherence. Equation~\eqref{eq:classical} is consistent with the Maxwell--Bloch equation derived from mean-field theory for the TC Hamiltonian and can be used to describe the probability flow in the spin subspace $(S_x,S_y,S_z)$.

\subsection{Spin trajectory}

We now consider a general solution to Eq.~\eqref{eq:classical} under the initial condition $\vec{S}(0)=S(\sin(\theta_0)\cos(\phi_0),\sin(\theta_0)\sin(\phi_0),-\cos(\theta_0))$, $ \vec\alpha(0)=(x_0,y_0)$. Because the Bloch-vector length $S$ is conserved, we use the spherical coordinates $(\theta,\phi)$ to describe the trajectory, where $\theta$ is the angle between the Bloch vector and the $-z$ axis. Substituting these expressions into Eq.~\eqref{eq:classical}, we obtain a closed set of equations for $\theta$ and $\phi$. In terms of the original time $t$, the three equations are
\begin{align}
\frac{d\alpha^*}{dt} &= igS\sin\theta\, e^{i\phi}, \tag{a}\\
i\frac{d}{dt}\left(S\sin\theta\, e^{i\phi}\right) &= -2gS\cos\theta\, \alpha^*, \tag{b}\\
\frac{d}{dt}\left(-S\cos\theta\right) &= 2g\,\Im\left[\alpha S\sin\theta\, e^{i\phi}\right]. \tag{c}
\end{align}
From Eq.~(b), we obtain $\alpha^* = -\frac{i}{2g\cos\theta}\frac{d}{dt}\left(\sin\theta\, e^{i\phi}\right)$, and substituting it into Eq.~(a) gives
$-\frac{d}{dt}\left[\frac{1}{\cos\theta}\frac{d}{dt}\left(\sin\theta\, e^{i\phi}\right)\right]
= 2g^2 S \sin\theta\, e^{i\phi}$.
Expanding the left-hand side and separating the real and imaginary parts gives
\begin{align}
\ddot{\theta} - \tan\theta\,\dot{\phi}^2 + 2g^2 S\sin\theta &= 0,\nonumber\\
\ddot{\phi} + \cot\theta\left(1+\frac{1}{\cos^2\theta}\right)\dot{\theta}\dot{\phi} &= 0.\label{eq:theta}
\end{align}

By introducing the dimensionless time $\tau=g\sqrt{2S}t$ and performing the first integration, one obtains
\begin{equation}
\begin{split}
&\phi'\frac{\sin^2(\theta)}{\cos(\theta)}=L\\
&\theta'^2 +V(\theta)=E,
\label{eq:first_integrals}
\end{split}
\end{equation}
where $V(\theta)=\frac{L^2}{\sin^2(\theta)}-2\cos(\theta)$ is the effective potential for the $\theta$ motion. Both $L$ and $E$ are integration constants determined by the initial conditions: $L=\frac{2\sin(\theta_0)}{\sqrt{2S}}(\cos(\phi_0 )x_0-\sin(\phi_0)y_0)$ and $E=\frac{2}{S}(x_0^2+y_0^2)-2\cos(\theta_0 )$. The initial conditions are $\theta(0)=\theta_0$, $\theta'(0)=\frac{2}{\sqrt{2S}}(\sin(\phi_0)x_0+\cos(\phi_0)y_0)$, $\phi(0)=\phi_0 $, and $\phi'(0)=\frac{2\cot(\theta_0)}{\sqrt{2S}}(\cos(\phi_0 )x_0-\sin(\phi_0)y_0)$.

From Eq.~\eqref{eq:first_integrals}, one can identify that the motion of a single spin trajectory is analogous to that of a fixed-length pendulum in a uniform gravitational field. The conserved quantity $L$ is the pseudo-angular momentum, and $E$ is the pseudo-energy related to the total excitation number $N_e$ of the TC model with $N_e=|\alpha(0)|^2-S\cos(\theta_0)$.

\begin{figure*}[t]
\includegraphics[width=\textwidth]{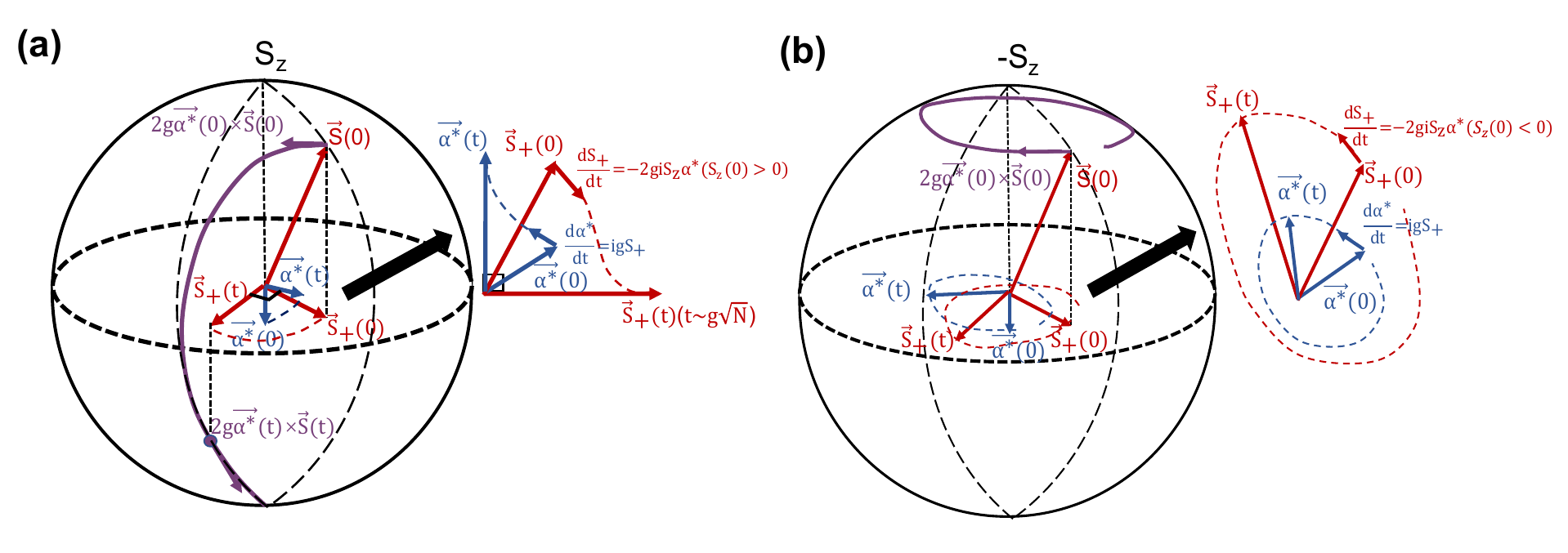}
\caption{\textbf{Vacuum fluctuation contribution to the spin-field trajectories under small and large spin excitations.}
Panels~(a) and~(b) show semiclassical trajectories of the collective spin
$\vec{S}(t)$ and cavity field $\alpha^*(t)$ governed by
$\dot{\alpha}^*=igS_+$ and $\dot{\vec{S}}=2g\,\vec{\alpha}^*\times\vec{S}$.
In panel~(a), for initial states near the excited state [$S_z(0)>0$], the radiated field
rapidly dominates over fluctuations, producing strong phase locking and Rabi
oscillations along a fixed-phase arc.
In panel~(b), for initial states near the ground state [$S_z(0)<0$], the radiated field is
comparable to vacuum fluctuations, leading to weak phase locking and
fluctuation-driven conical motion on the Bloch sphere.}
\label{fig:supp_fig1}
\end{figure*}

\subsection{Vacuum fluctuations}

We now consider the characteristics of the spin trajectory. For an initial condition without vacuum fluctuations, $\alpha(0)=0$, or satisfying $\arg(iS_+)=\arg(\alpha^*)+k\pi,k\in Z$, one has $\phi'=0$, and Eq.~\eqref{eq:theta} becomes $\ddot{\theta}+g^2N\sin(\theta)=0$ in the large-$N$ limit. The trajectory is therefore equivalent to that of a dissipationless pendulum moving along a meridian passing through the south pole.

To include vacuum-fluctuation effects, the light-field quadratures $x_0$ and $y_0$ can be treated as random variables centered at zero with variance $1/4$. This variance is much smaller than $\sqrt{N}$, so that $L\ll1$. The trajectory depends on the initial spin orientation $\theta_0$, because the total light field emitted by the spin ensemble is bounded from above by $\sqrt{N}\sin(\theta_0/2)$ and tends to ``lock'' the spin trajectory to a meridian. We discuss the underlying physical mechanism below.

In a superradiant-emission process, the optical field generated by spontaneous emission acquires a phase shifted by $\pi/2$ relative to the spin coherence, as expressed by $\dot{\alpha^*}=igS_+$. The spin ensemble subsequently undergoes Rabi dynamics driven by this self-generated field. Its rotation axis is aligned with the phase of the optical coherence, and the dynamics can be summarized as $\dot{\vec{S}}=2g\vec{\alpha^*}\times\vec{S}$, where $\vec{\alpha^*}=(\Re(\alpha^*),\Im(\alpha^*))$ is the light-coherence vector in the $x$--$y$ plane. When the initial optical field is the vacuum state without fluctuations, all intracavity photons originate from spin radiation. Under a driving field whose phase is displaced by $\pi/2$ from the spin coherence, the collective spin undergoes Rabi oscillations along an arc on the Bloch sphere, while the spin phase remains constant. The optical phase is likewise conserved. Because photon emission and reabsorption accompany the motion, the instantaneous Rabi frequency is not constant, and the resulting dynamics resemble a pendulum-like back-and-forth motion along the arc.

Including vacuum fluctuations modifies this picture. The initial optical field acquires a coherence of the order of the fluctuation amplitude, which induces an initial Rabi rotation of the spin. Meanwhile, the spin radiates an optical field whose phase is orthogonal to its own coherence phase, thereby favoring phase-preserving Rabi motion. For initial states near the excited state (north pole), as shown in Fig.~\ref{fig:supp_fig1}(a), the radiated field rapidly exceeds the fluctuation amplitude and dominates the intracavity field. The resulting phase locking between radiation and Rabi dynamics drives the spin toward a well-defined phase. For initial states near the ground state, the radiated field is comparable to the vacuum fluctuations, and the phase-locking mechanism becomes weak. The spin trajectory then exhibits fluctuation-driven, conical-pendulum-like motion, as shown in Fig.~\ref{fig:supp_fig1}(b).

\begin{figure*}[t]
\includegraphics[width=\textwidth]{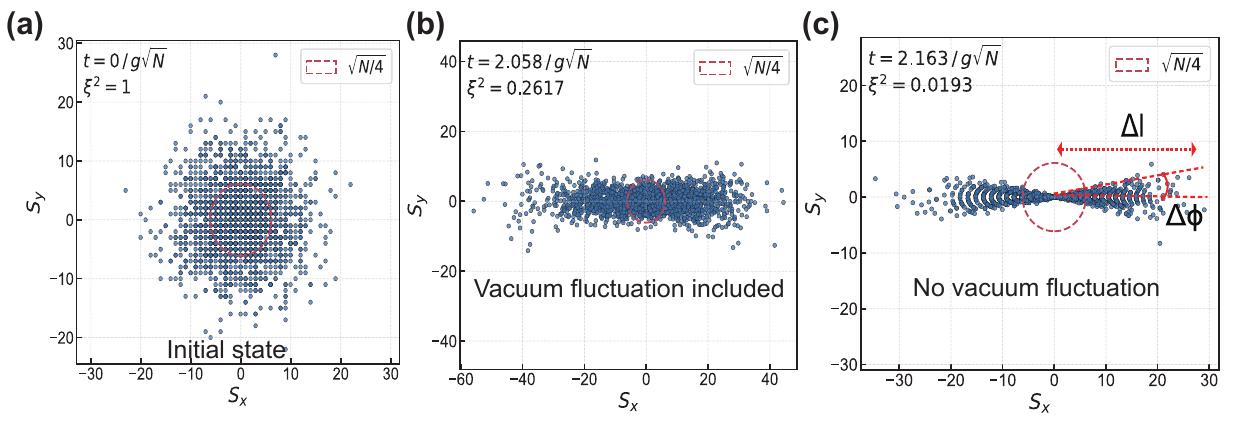}
\caption{\textbf{Spin Wigner distributions.}
(a) Wigner function of the initial CSS for $N=150,\theta_0=120.6^{\circ}$, showing a homogeneous Gaussian distribution with variance $C_{xx}=C_{yy}=N/4$.
(b) Spin Wigner distribution of the optimal squeezed state near the south pole obtained using the complete DTWA method from the initial state shown in panel~(a), with squeezing $\xi^2=0.26$ at $t=2.058/g\sqrt{N}$.
(c) Spin Wigner distribution of the optimal squeezed state near the south pole obtained using DTWA from the initial CSS in panel~(a), without vacuum fluctuations in the initial cavity mode. The distribution has an ``$\infty$-shaped'' non-Gaussian form, with squeezing $\xi^2=0.0193$ at $t=2.163/g\sqrt{N}$. The geometric parameters $\Delta \phi$ and $\Delta l$ can be used to estimate the optimal squeezing parameter.
All panels are obtained by rotating the mean-spin direction to the $+z$ axis and projecting the distribution onto the $S_x$--$S_y$ plane. A reference circle with radius $\sqrt{N/4}$ is also shown.}
\label{fig:wigner}
\end{figure*}

Within the DTWA framework, collective-spin correlation functions are obtained by sampling the initial Wigner distribution and averaging over the resulting classical trajectories. In the absence of vacuum fluctuations, all trajectories become phase locked, evolve along the meridian defined by the initial phase, and converge at the south pole. The Wigner quasiprobability distribution of the optimal squeezed state near the south pole is shown in Fig.~\ref{fig:wigner}(c), where it has an ``$\infty$''-shaped form. The initial CSS for $N=150, \theta_0=120.6^{\circ}$ is shown in Fig.~\ref{fig:wigner}(a). After vacuum fluctuations are included, the distribution becomes approximately Gaussian, as shown in Fig.~\ref{fig:wigner}(b), and the attainable squeezing is reduced.

\subsection{Analysis of the squeezing parameter}

As shown in Fig.~\ref{fig:wigner}, vacuum fluctuations reduce the attainable squeezing. Estimating the optimal squeezing parameter therefore requires a detailed analysis of the competition between vacuum fluctuations and the main squeezing mechanism, namely, the geometric focusing effect induced by the curvature of the Bloch sphere.

To estimate the optimal squeezing parameter in the absence of vacuum fluctuations, we use the width $\Delta\phi$ of the spin noise and the effective length $\Delta l$ along the meridian to describe the distribution. The quantity $\Delta\phi=\arcsin\left(\frac{\arcsin(1/\sqrt{N})}{\sin(\theta_0)}\right)\approx\frac{1}{\sqrt{N}\sin(\theta_0)}$ is determined by the initial CSS and is conserved during the evolution. The quantity $\Delta l$ can be calculated from the stretching along the meridian caused by the superradiant delay, $\Delta l\approx\frac{\sqrt{N}}{2} \frac{\partial\theta}{\partial\theta_0}|_{t=t_{sq}}$. The resulting dependence of the optimal squeezing parameter on $\theta_0$ and $N$ is
\begin{equation}
    \xi^2(\theta_0,N)\approx N\frac{(\Delta\phi\Delta l)^2}{S^2}\approx \frac{1}{
    N\sin^2\theta_0}\left(\frac{E\left(\sin\frac{\theta_0}{2}\right)}{\cos(\frac{\theta_0}{2})}-\cos\left(\frac{\theta_0}{2}\right)K\left(\sin\frac{\theta_0}{2}\right)\right)^2,
\label{eq:squeezing_no_vacuum}
\end{equation}
where $K,E$ are the complete elliptic integrals of the first and second kinds, respectively.

\begin{figure*}[t]
\includegraphics[width=0.9\textwidth]{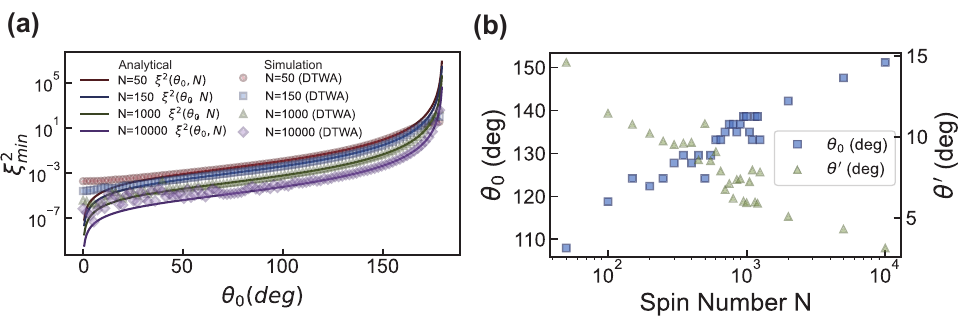}
\caption{
(a) Dependence of the optimal squeezing parameter near the south pole on the initial angle $\theta_0$ and spin number $N$ in the absence of vacuum fluctuations. The analytical result $\xi^2(\theta_0,N)$ is compared with DTWA results for $N=50,150,1000,10000$. Good agreement is obtained over most of the $\theta_0$ range, except near $0^{\circ}$ and $180^{\circ}$, where the approximation breaks down.
(b)  Optimal initial angle $\theta_0$ and corresponding squeezing position $\theta'$ for different spin numbers $N$. The optimal initial angle lies in the interval $[90^{\circ},180^{\circ}]$ and approaches $180^{\circ}$ monotonically as $N$ increases, while $\theta'$ approaches zero. }
\label{fig:supp_fig3}
\end{figure*}

We consider the time evolution of the semimajor and semiminor axes of the noise ellipse to estimate the squeezing parameter~\cite{wineland_squeezed_1994} along the corresponding directions. We define $\xi^2_1$ as the squeezing parameter along $Y$ (parallel) and $\xi^2_2$ as that along $X$ (meridional). The two parameters can be approximated as
\begin{equation}
\begin{split}
\xi^2_1(t,\theta_0,N) &= \left(\frac{\sin(\theta(t))}{\sin(\theta_0)}\right)^2, \\
\xi^2_2(t,\theta_0,N) &= \left(\frac{\partial\theta(t)}{\partial \theta_0}\right)^2,
\end{split}
\label{eq:xi}
\end{equation}
where $
\theta(\theta_0,t) =
2\arcsin\left(
\sin\left(\frac{\theta_0}{2}\right)
\operatorname{sn}\left(
K\left(\sin\left(\frac{\theta_0}{2}\right)\right) - g\sqrt{N}t,\;
\sin\left(\frac{\theta_0}{2}\right)
\right)
\right)
$
is the trajectory starting from the initial position $\theta_0$, and $\text{sn}$ is the Jacobi elliptic sine function. The approximate time of the first squeezing event is obtained by setting $\xi^2_1=0$,
\begin{equation}
    T_{squeezing}= \frac{K\left(\sin\left(\frac{\theta_0}{2}\right)\right)}{g\sqrt{N}}.
    \label{T_sq}
\end{equation}

In the absence of vacuum fluctuations, smaller $\theta_0$ is preferred for stronger squeezing, as shown in Fig.~\ref{fig:supp_fig3}(a). However, vacuum fluctuations play a competing role and favor larger $\theta_0$, because a stronger radiated field suppresses the expansion of the spin variance along the squeezing direction, as discussed in the previous subsection. The simulated optimal initial angle $\theta_0$ and squeezing position $\theta'$ as functions of $N$ are shown in Fig.~\ref{fig:supp_fig3}(b). The optimal initial angle lies in the northern hemisphere and approaches $180^{\circ}$ as $N$ increases, while the corresponding squeezing position $\theta^\prime$ approaches zero.

\subsection{Stochastic phase-space description}

In the presence of dissipation, the noise should be added to the evolution of one trajectory. For simplicity, we assume that the noise is Markovian white noise, and that the interaction between the system and the environment can be described by Lindblad operators. For each emitter, we introduce classical spin variables
$s_k^{x}$, $s_k^{y}$, and $s_k^{z}$ corresponding to the Pauli operators $\hat{\sigma}_k^x
    =
    \hat{\sigma}_k^{12}
    +
    \hat{\sigma}_k^{21},
    \hat{\sigma}_k^y
    =
    -i
    \left(
        \hat{\sigma}_k^{12}
        -
        \hat{\sigma}_k^{21}
    \right),
    \hat{\sigma}_k^z
    =
    \hat{\sigma}_k^{22}
    -
    \hat{\sigma}_k^{11}.$

The initial spin state is sampled using a discrete Wigner representation. For each spin $k$, the allowed phase-space points are $\mathbf{s}_k
    =
    p_1\mathbf{s}_1
    +
    p_2\mathbf{s}_2
    +
    p_3\mathbf{s}_3,
    p_1,p_2,p_3=\pm 1,$
with the local basis vectors $\mathbf{s}_1
    =
    \left(
        \cos\theta_0,0,\sin\theta_0
    \right), 
    \mathbf{s}_2
    =
    \left(
        0,1,0
    \right),
    \mathbf{s}_3
    =
    \left(
        -\sin\theta_0,0,\cos\theta_0
    \right).$
The corresponding single-spin Wigner distribution is
\begin{equation}
    W_{0k}
    \left(
        p_{1k},p_{2k},p_{3k}
    \right)
    =
    \frac{1}{8}
    \left(
        1-p_{3k}
    \right).
    \label{eq:single_spin_wigner}
\end{equation}
This distribution reproduces the Bloch vector of the initial spin-coherent state $\overline{\mathbf{s}_k}
    =
    \left(
        \sin\theta_0,0,-\cos\theta_0
    \right)$, where the overline denotes an average over the initial Wigner distribution.

The cavity field is represented by two real phase-space quadratures $x$ and $y$. For the vacuum state, the initial Wigner distribution is Gaussian, $W_{0}(x,y)
    =
    \frac{2}{\pi}
    \exp
    \left[
        -2
        \left(
            x^2+y^2
        \right)
    \right].$ Thus, the full initial Wigner function factorizes as
\begin{equation}
    W
    \left(
        \{s_k^\alpha\},x,y
    \right)
    =
    \prod_{k=1}^{N}
    W_{0k}
    \left(
        p_{1k},p_{2k},p_{3k}
    \right)
    W_{0}(x,y),
    \label{eq:full_initial_wigner}
\end{equation}
where $\alpha=x,y,z$. For each stochastic trajectory, the classical field and spin variables evolve according to the following stochastic differential equations~\cite{hosseinabadi_userfriendly_2025}:
\begin{subequations}
\label{eq:stochastic_equations}
\begin{align}
    \frac{dx}{dt}
    &=
    \omega_{c} y
    -
    \frac{g}{2}
    \sum_{k=1}^{N}s_k^y
    -
    \frac{\kappa}{2}x
    -
    \frac{1}{2}\xi_{\kappa,1},
    \label{eq:dxdt}
    \\
    \frac{dy}{dt}
    &=
    -\omega_{c} x
    -
    \frac{g}{2}
    \sum_{k=1}^{N}s_k^x
    -
    \frac{\kappa}{2}y
    -
    \frac{1}{2}\xi_{\kappa,2},
    \label{eq:dydt}
    \\
    \frac{ds_k^x}{dt}
    &=
    -\omega_{a} s_k^y
    -
    2gs_k^z y
    +
    \frac{\Gamma}{2}
    s_k^x s_k^z
    +
    \xi_{\downarrow,k}^{x}
    s_k^z,
    \label{eq:dsxdt}
    \\
    \frac{ds_k^y}{dt}
    &=
    \omega_a{} s_k^x
    -
    2gs_k^z x
    +
    \frac{\Gamma}{2}
    s_k^y s_k^z
    +
    \xi_{\downarrow,k}^{y}
    s_k^z,
    \label{eq:dsydt}
    \\
    \frac{ds_k^z}{dt}
    &=
    2g s_k^y x
    +
    2g s_k^x y
    -
    \frac{\Gamma}{2}
    \left[
        \left(s_k^x\right)^2
        +
        \left(s_k^y\right)^2
    \right]
    \nonumber \\
    &\quad
    -
        \xi_{\downarrow,k}^{x}
    s_k^x
    -
        \xi_{\downarrow,k}^{y}
s_k^y .
    \label{eq:dszdt}
\end{align}
\end{subequations}
Here $\kappa$ denotes the cavity decay rate, while $\Gamma$ describes the spin relaxation (spontaneous emission). The stochastic fields are Gaussian white noises with zero mean and correlations
\begin{subequations}
\label{eq:noise_correlations}
\begin{align}
    \overline{
        \xi_{\kappa,\alpha}(t)
        \xi_{\kappa,\beta}(t')
    }
    &=
    \kappa
    \delta_{\alpha\beta}
    \delta(t-t'),
    \qquad
    \alpha,\beta=1,2,
    \label{eq:kappa_noise}
    \\
    \overline{
        \xi_{\downarrow,i}^{\alpha}(t)
        \xi_{\downarrow,j}^{\beta}(t')
    }
    &=
    \Gamma
    \delta_{\alpha\beta}
    \delta_{ij}
    \delta(t-t'),
    \qquad
    \alpha,\beta=x,y,
    \label{eq:down_noise}
\end{align}
\end{subequations}
All remaining noise correlations vanish.

Physical observables are obtained by averaging over an ensemble of $N_{\mathrm{tr}}$ independent stochastic trajectories. For an arbitrary phase-space observable $O$, the stochastic estimate is
\begin{equation}
    \langle \hat{O}(t)\rangle
    \simeq
    \frac{1}{N_{\mathrm{tr}}}
    \sum_{r=1}^{N_{\mathrm{tr}}}
    O^{(r)}(t),
    \label{eq:trajectory_average}
\end{equation}
\begin{figure*}[t]
\centering
\includegraphics[width=0.5\textwidth]{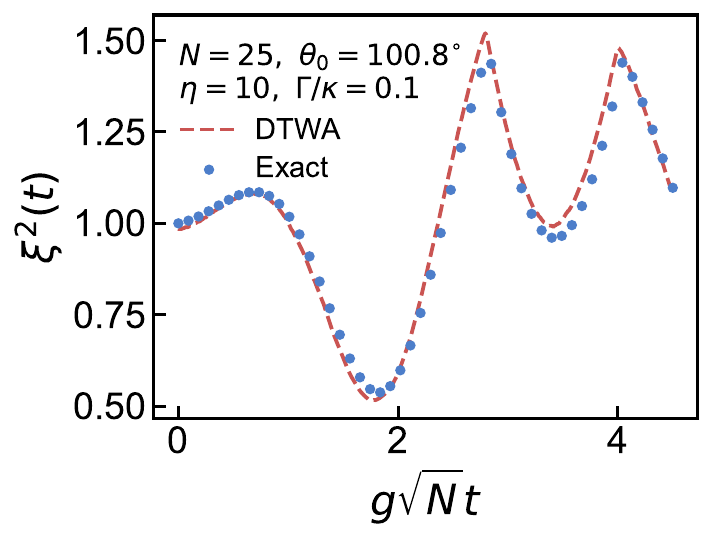}
\caption{Comparison of the evolution of squeezing parameter $\xi^2(t)$ between DTWA method and the exact solution. In the diagram we take $N=25, \theta_0=100.8^{\circ}$ and $\eta=10, \Gamma/\kappa=0.1$ for example, which shows the consistency between the two solutions.}
\label{xi_comparison}
\end{figure*}
where $O^{(r)}(t)$ is evaluated along the $r$th trajectory. The initial conditions of each trajectory are sampled independently from the factorized Wigner distribution in Eq.~\eqref{eq:full_initial_wigner}. This stochastic phase-space formulation therefore incorporates the initial quantum fluctuations of both the cavity vacuum and the spin-coherent state, while the subsequent dynamics is propagated at the level of classical stochastic equations, the consistency compared to the exact solution of the master equation is shown in Fig.~\ref{xi_comparison}.

\section{Light-mode squeezing}

From Eq.~\eqref{eq:classical}, one can derive the equation of motion for the light field,
\begin{equation}
\frac{d^2\alpha^*}{dt^2}+2g^2(|\alpha|^2-N_e)\alpha^*=0,
\label{eq:light_squeezing}
\end{equation}
where $N_e=|\alpha(0)|^2-S\cos(\theta_0)$ is the conserved excitation number. For a specific trajectory, the initial conditions are $\alpha^*(0)=\alpha_0^*$ and $\dot\alpha^*(0)=igS\sin(\theta_0)e^{i\phi_0}$.

\begin{figure*}[t]
\centering
\includegraphics[width=\textwidth]{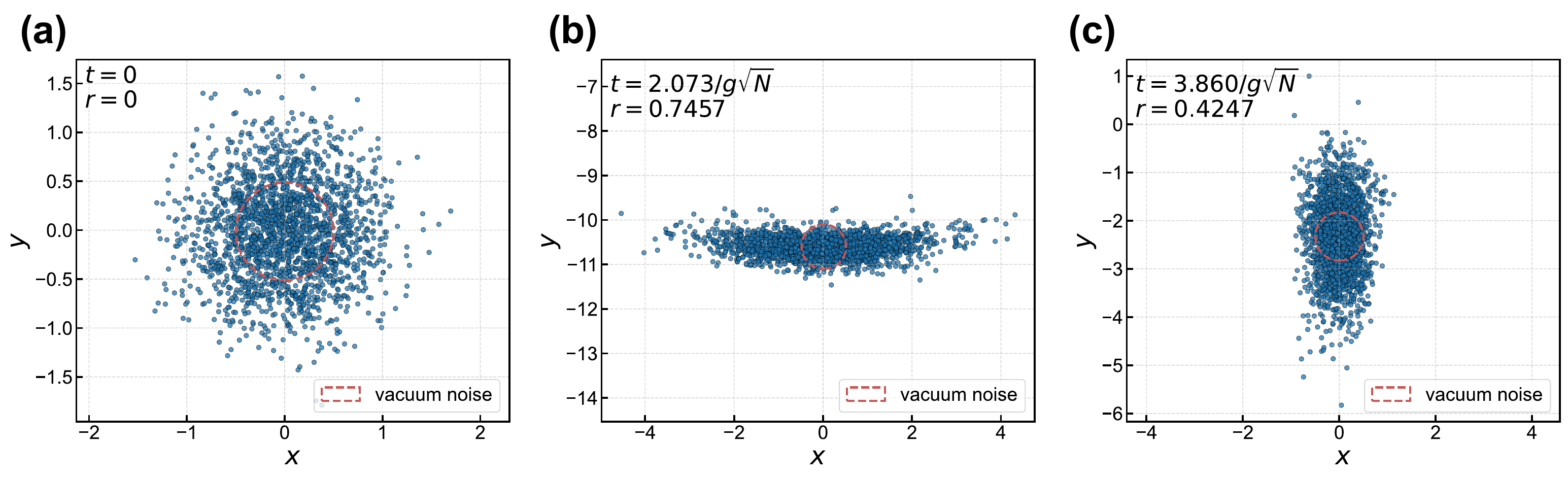}
\caption{\textbf{Wigner distributions of the squeezed light mode accompanying the squeezed spin states.}
(a) Wigner distribution of the initial vacuum state, with isotropic variance $1/4$.
(b) Squeezed light-mode Wigner distribution accompanying the first spin-squeezing event. Squeezing occurs along the $y$ direction, with variance $\Delta Y^2=\frac{1}{4}e^{-2r}$ and $r=0.7457$. The state contains multiple photons and has light coherence $\Im(\alpha)=-\sqrt{N}\sin\left(\theta_0/2\right)=-10.6$.
(c) Squeezed light-mode Wigner distribution accompanying the second spin-squeezing event. Squeezing occurs along the $x$ direction, with variance $\Delta X^2=\frac{1}{4}e^{-2r}$ and $r=0.4247$. Most photons have been reabsorbed, and the light coherence is close to zero.
The parameters are $N=150,\theta_0=120.6^{\circ}$. A circle corresponding to the vacuum-noise standard deviation is shown for reference.}
\label{fig:supp4_new}
\end{figure*}

During vacuum Rabi oscillations, the cavity field fluctuations become squeezed along the $x$ [$\Re(\alpha)$] or $y$ [$\Im(\alpha)$] quadrature in conjunction with the spin-squeezing dynamics. The Wigner distributions of the light mode are shown in Fig.~\ref{fig:supp4_new}. As the light field oscillates along the $y$ axis, squeezing occurs sequentially along the $y$ and $x$ directions. Similar to spin squeezing, light squeezing can be understood intuitively by neglecting the initial spin noise. Any trajectory starting from $(x_0,y_0)$ then has the same initial velocity $\dot x=0$, $\dot y=-gS\sin(\theta_0)$.

Equation~\eqref{eq:light_squeezing} contains the nonlinear term $2g^2|\alpha|^2\alpha^*$, whose solution depends sensitively on the initial light field $\alpha_0$. The nonuniform photon-oscillation rate causes light squeezing. For the trajectory whose initial coherence is located at $(0,0)$, the solution is $\alpha(t)=i\sqrt{N}\sin\left(\frac{\theta_0}{2}\right)\text{cn}\left(g\sqrt{N}t+K(\sin\left(\frac{\theta_0}{2}\right)\right),\sin\left(\frac{\theta_0}{2}\right))$,
where cn and $K$ denote the Jacobi elliptic cosine function and the complete elliptic integral of the first kind, respectively. The squeezing parameter can also be estimated by calculating the line elements of the standard deviations along the $x$ and $y$ directions. For trajectories initially located at $(0,\Delta Y_0)$ and $(\Delta X_0,0)$, where $\Delta X_0=\Delta Y_0=1/2\ll\sqrt{N}$, the line-element evolution satisfies the standard $\text{Lam\'e}$ equation for $n=1$ ($\Delta X$) and $n=2$ ($\Delta Y$)~\cite{maier_lame_2008}.

\begin{equation}
\begin{split}
\Delta X(t,\theta_0) &= \frac{\Delta X_0}{\sqrt{1-k^2}} \Bigg[
\operatorname{sn}(g\sqrt{N}t+K(k),k)\,\operatorname{dn}(g\sqrt{N}t+K(k),k) \\
&\quad + \operatorname{cn}(g\sqrt{N}t+K(k),k) \Big( E(K(k)) - E(g\sqrt{N}t+K(k)) + (1-k^2)g\sqrt{N}t \Big) \Bigg], \\
\Delta Y(t,\theta_0) &= \frac{\Delta Y_0}{\sqrt{1 - k^2}}
\operatorname{sn}(g\sqrt{N}t+K(k),k)\,\operatorname{dn}(g\sqrt{N}t+K(k),k).
\end{split}
\label{eq:10}
\end{equation}
Here, sn, cn, and dn are Jacobi elliptic functions, while $E$ and $K$ are complete elliptic integrals. The modulus is defined as $k=\sin\left(\frac{\theta_0}{2}\right)$. The squeezing parameter~\cite{yuen_twophoton_1976} is $r_i(t,\theta_0)=-1/2\ln\left(\frac{\Delta X_{i}(t,\theta_0)}{\Delta X_{i0}}\right)^2$ for $i=x,y$. 

By analyzing the zero of $r(t,\theta_0)$, one can approximate the squeezing time by setting $\Delta Y=0$,
\begin{equation}T_{squeezing}=\frac{K\left(\sin\left(\frac{\theta_0}{2}\right)\right)}{g\sqrt{N}},
\end{equation}
which is consistent with Eq.~\eqref{T_sq} and demonstrates simultaneous occurrence of spin and light-mode squeezing during vacuum Rabi oscillations.

\section{Cumulant-expansion description}

We employ mean-field theory~\cite{plankensteiner_quantumcumulantsjl_2022} to analyze the spin-squeezing dynamics accompanying vacuum Rabi oscillations. In the resonant strong-coupling regime, the master equation~\cite{ tavis_exact_1968,stitely_quantum_2023}for the density matrix is
\begin{equation}
    \frac{d\hat{\rho}}{dt}=-\frac{i}{\hbar}[\hat{H}_{TC},\hat{\rho}]-\frac{\Gamma}{2}\sum_{i}^{N}\left(\hat{\rho}\sigma_i^{\uparrow\uparrow}+\sigma_i^{\uparrow\uparrow}\hat{\rho}-2\sigma_i^{\downarrow\uparrow}\hat{\rho}\sigma_i^{\uparrow \downarrow}\right)-\frac{\kappa}{2}\left(\hat{\rho}\hat{a}^{\dagger}\hat{a}+\hat{a}^{\dagger}\hat{a}\hat{\rho}-2\hat{a}\hat{\rho}\hat{a}^{\dagger}\right),
    \label{master}
\end{equation}
where the spin operator is defined as $\sigma_i^{\downarrow\uparrow}=(\sigma_i^{\uparrow\downarrow})^{\dagger}=|\downarrow_i\rangle\langle\uparrow_i|$, and the Tavis--Cummings Hamiltonian is $\hat{H}_{TC}=\hbar g(\hat{a}\hat{S}_++\hat{a}^{\dagger}\hat{S}_-)$. To capture the spin-noise dynamics correctly, the cumulant expansion must be carried out to at least second order so that genuine quantum effects are retained.

For the initial state $|CSS\rangle\otimes|0\rangle$ under resonant conditions, we define the spin and light coherences as $s_i=\langle S_i\rangle$ and $(x,y)=(\langle \frac{a+a^{\dagger}}{2}\rangle,\langle \frac{a-a^{\dagger}}{2i}\rangle)$. The spin covariance is defined as $C_{ij}=\langle \frac{S_iS_j+S_jS_i}{2}\rangle-\langle S_i\rangle\langle S_j\rangle$ for $i,j=x,y,z$. The light covariance is defined as $C_{\alpha\beta}=\langle \frac{X_\alpha X_{\beta}+X_{\beta}X_{\alpha}}{2}\rangle-\langle X_{\alpha}\rangle\langle X_{\beta}\rangle$ for $\alpha,\beta=1,2$, where $X_1=X=\frac{a+a^{\dagger}}{2}$ and $X_2=Y=\frac{a-a^{\dagger}}{2i}$. Finally, the spin--light correlation is defined as $C_{\alpha i}=\langle S_i X_\alpha\rangle-\langle S_i\rangle \langle X_{\alpha}\rangle$ for $i=x,y,z$ and $\alpha=1,2$. The complete second-order mean-field equations derived from Eq.~\eqref{master} are
\begin{equation}
\begin{aligned}
\frac{d s_x}{dt}
&=-2gys_z-2gC_{2z}-\frac{\Gamma}{2}s_x,\\
\frac{d s_z}{dt}
&=2gys_x+2gC_{2x}+2gC_{1y}
-\Gamma s_z-\frac{\Gamma N}{2},\\
\frac{d y}{dt}
&=-gs_x-\frac{\kappa}{2}y,\\
\frac{d C_{xx}}{dt}
&=-4gyC_{xz}-4gs_zC_{2x}
-\Gamma C_{xx}+\frac{\Gamma N}{4},\\
\frac{d C_{yy}}{dt}
&=-4gs_zC_{1y}
-\Gamma C_{yy}+\frac{\Gamma N}{4},\\
\frac{d C_{zz}}{dt}
&=4gyC_{xz}+4gs_xC_{2z}
-2\Gamma C_{zz}+\Gamma s_z+\frac{\Gamma N}{2},\\
\frac{d C_{xz}}{dt}
&=2g\left(
yC_{xx}+s_xC_{2x}-yC_{zz}-s_zC_{2z}
\right)
+\frac{\Gamma}{2}s_x-\frac{3\Gamma}{2}C_{xz},\\
\frac{d C_{2x}}{dt}
&=-2gyC_{2z}-2gs_zC_{22}
-gC_{xx}-\frac{\Gamma+\kappa}{2}C_{2x},\\
\frac{d C_{1y}}{dt}
&=-2gs_zC_{11}-gC_{yy}
-\frac{\Gamma+\kappa}{2}C_{1y},\\
\frac{d C_{2z}}{dt}
&=2gyC_{2x}+2gs_xC_{22}
-gC_{xz}-\frac{\Gamma+\kappa}{2}C_{2z},\\
\frac{d C_{11}}{dt}
&=-2gC_{1y}-\kappa C_{11}+\frac{\kappa}{4},\\
\frac{d C_{22}}{dt}
&=-2gC_{2x}-\kappa C_{22}+\frac{\kappa}{4},
\label{CE}
\end{aligned}
\end{equation}
with the symmetry-protected quantities $C_{xy}=C_{yz}=C_{1x}=C_{2y}=C_{1z}=C_{12}=s_y=x=0$.

\begin{figure*}[t]
\centering
\includegraphics[width=\textwidth]{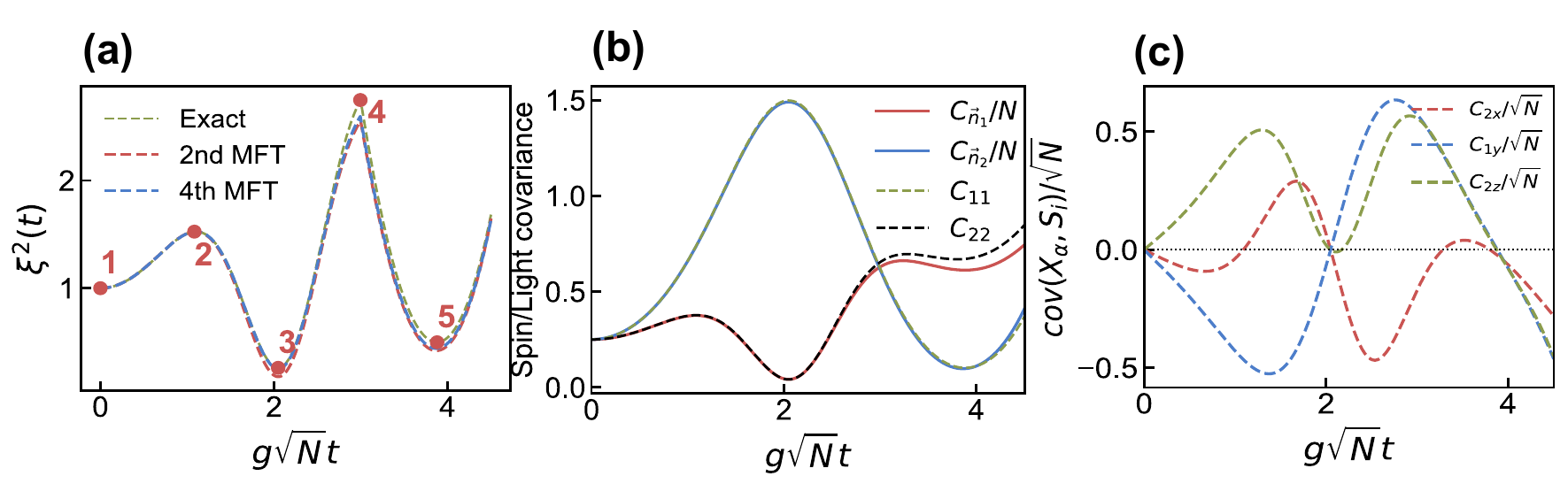}
\caption{
(a) Comparison of the squeezing-parameter evolution obtained from the exact solution and from second- and fourth-order mean-field theories. The results are mutually consistent.
(b) Evolution of the spin and light eigen-variances obtained from second-order mean-field theory. Spin noise is transferred to the light mode: squeezing of $C_{\vec{n}_1}$ is accompanied by anti-squeezing of $C_{11}$, while squeezing of $C_{\vec{n}_2}$ is accompanied by anti-squeezing of $C_{22}$. The transfer rate is proportional to the spin--light correlation. In the large-$N$ limit, the curves of $C_{11}$ and $C_{\vec{n}_2}/N$ coincide, as do those of $C_{22}$ and $C_{\vec{n}_1}/N$.
(c) Evolution of the spin--light correlations obtained from second-order mean-field theory. The correlations are proportional to the noise-transfer rates, scale as $\sqrt{N}$, and approach zero at the two optimal squeezing points. This behavior indicates weak residual spin-light correlations at the optimal squeezing points, while both subsystems retain strongly nonclassical squeezed fluctuations. The parameters are $N=150,\theta_0=120.6^{\circ}$.}
\label{supp5}
\end{figure*}

To identify squeezing, we define two directions perpendicular to the unit Bloch vector $\vec{n}_3=(\sin(\theta),0,-\cos(\theta))$: 
$\vec{n}_1=(0,1,0)$ and 
$\vec{n}_2=\frac{\vec{n}_3\times\vec{n}_1}{|\vec{n}_3\times\vec{n}_1|}$. The spin covariance matrix in the $\vec{n}_1,\vec{n}_2$ plane is
\begin{equation}
    \label{eq:matrix}
    V_p=
    \begin{bmatrix}
        C_{yy} & 0 \\
        0 & \cos^2(\theta)C_{xx}+\sin^2(\theta)C_{zz}+2\sin(\theta)\cos(\theta)C_{xz}
    \end{bmatrix},
\end{equation}
which is diagonal, showing that $\vec{n}_1$ and $\vec{n}_2$ are eigen-directions of the spin noise. For the initial state, $C_{yy}=N/4,C_{xx}=(N/4)\cos^2(\theta_0),C_{zz}=(N/4)\sin^2(\theta_0),C_{xz}=(N/4)\sin(\theta_0)\cos(\theta_0)$. The covariance is therefore initially isotropic and equal to $N/4$. During the evolution governed by Eq.~\eqref{CE}, however, the variances along $\vec{n}_1$ and $\vec{n}_2$ oscillate and can fall below $N/4$. A comparison of the squeezing parameter $\xi^2$ obtained from the exact solution and from second- and fourth-order cumulant expansion theories is shown in Fig.~\ref{supp5}(a).

For the spin noise along the $y$ direction, the relevant equations are $\frac{d C_{yy}}{dt}
=-4gs_zC_{1y}
-\Gamma C_{yy}+\frac{\Gamma N}{4}$, $\frac{d C_{1y}}{dt}
=-2gs_zC_{11}-gC_{yy}
-\frac{\Gamma+\kappa}{2}C_{1y}$, and $\frac{d C_{11}}{dt}
=-2gC_{1y}-\kappa C_{11}+\frac{\kappa}{4}$. If the initial condition satisfies $s_z(0)>0$, then $dC_{yy}/dt>0$ initially for $C_{1y}<0,C_{11}>0$, and the variance along $y$ is anti-squeezed. When $s_z<0$ (after passing the equator), one has $dC_{yy}/dt<0$, corresponding to squeezing. This behavior is consistent with the geometric focusing caused by the curvature of the Bloch sphere.

For an initial angle $\theta_0$ sufficiently far from the north pole, dimensional analysis gives $C_{\alpha i}\sim\sqrt{N}$, which is much smaller than $s_i$ in the large-$N$ limit. By neglecting second-order correlations, the Bloch vector approximately satisfies the Bloch equation in the large-$N$ limit, with $\vec{s}=N/2(\sin(\theta),0,-\cos(\theta))$, and $\theta(t)$ is the same as in Eq.~\eqref{eq:xi}. The spin--light correlation $C_{1y}$ satisfies
\begin{equation}
    \dot{s}_z\,\dddot{C}_{1y}
    -\ddot{s}_z\,\ddot{C}_{1y}
    -8g^2s_z\dot{s}_z\,\dot{C}_{1y}
    +\left(
        8g^2s_z\ddot{s}_z
        -12g^2\dot{s}_z^{\,2}
    \right)C_{1y}
    =0
    \label{C1y}
\end{equation}
with the initial conditions $C_{1y}=0,\dot{C}_{1y}=-gN(1-\cos(\theta_0))/4,\ddot{C}_{1y}=0$. The quantities $C_{yy}$ and $C_{11}$ can be expressed as $C_{yy}=-4g\int_{0}^{t}s_zC_{1y}dt+N/4$ and $C_{11}=-2g\int_{0}^{t}C_{1y}dt+1/4$.

The equations for $C_{yy},C_{1y},C_{11}$ and the remaining equations involving $C_{\vec{n}_2},C_{22},C_{2x},C_{2z}$ reveal the close relationship between spin and light squeezing, where $C_{\vec{n}_1}, C_{\vec{n}_2}$ are the spin variances along $\vec{n}_1,\vec{n}_2$. In the absence of dissipation, $\frac{d C_{yy}}{dt}
=-4gs_zC_{1y}
,\frac{d C_{11}}{dt}
=-2gC_{1y}$ show that both the spin noise and light noise are influenced by  the spin--light correlation $C_{1y}$. When the spin noise $C_{yy}$ is squeezed ($\dot{C}_{yy}<0$) for $s_z<0$, the light noise $C_{11}$ is simultaneously anti-squeezed ($\dot{C}_{11}>0$) for $C_{1y}<0$. For the variance along $\vec{n}_2$, one can similarly derive $\frac{dC_{\vec{n}_2}}{dt}=-4gs_zC_{2x}+4gs_xC_{2z}$, which is related to the spin--light correlations $C_{2x},C_{2z}$.

The overall process can therefore be understood as noise transfer between the spin and light sectors induced by the cavity--spin interaction, at a rate proportional to the spin--light correlation. The transfer channels are $C_{\vec{n}_1}\longleftrightarrow C_{11}$ and $C_{\vec{n}_2}\longleftrightarrow C_{22}$. The first squeezing event corresponds to squeezing of $C_{\vec{n}_1}$ and $C_{22}$, while the second involves the other two directions. The optimal squeezing point occurs when the spin--light correlation changes sign, i.e., $cov(a,S_i)\approx0$ (Fig.~\ref{supp5}(c)), where pronounced squeezing remains in both the spin and light sectors. The dissipationless evolution for $N=150,\theta_0=120.6^{\circ}$ is shown in Fig.~\ref{supp5}(b).

We claim that, in the large-$N$ limit, the evolutions of the light noises $C_{11}$ and $C_{22}$ coincide with those of the normalized spin noises $C_{\vec{n}_2}/N$ and $C_{\vec{n}_1}/N$ respectively. For simplicity, we consider only $\Delta =C_{\vec{n}_1}/N-C_{22}$ to demonstrate that it vanishes as $N\xrightarrow{}\infty$. The evolution of $\Delta$ satisfies
\begin{equation}
    \frac{d\Delta}{dt}=2gC_{2x}+2g\cos(\theta(t))C_{1y}.
\end{equation}

The initial conditions are $\Delta(0)=\dot\Delta(0)=\ddot{\Delta}(0)=\dddot{\Delta}(0)=0,\ddddot{\Delta}(0)=Ng^4(1-\cos(\theta_0))^2(2\cos(\theta_0)-1)$. Hence, $\Delta\approx\frac{(1-\cos(\theta_0))^2(2\cos(\theta_0)-1)}{24N}(g\sqrt{N}t)^4+O((g\sqrt{N}t)^6)$. In the large-$N$ limit, $\Delta\xrightarrow{}0$ on the timescale $1/g\sqrt{N}$. Therefore, the evolution of the spin and light noise can be estimated from the solution of Eq.~\eqref{C1y}. The results are also consistent with simultaneous squeezing of the light and spin noises.

The exact solution to Eq.~\eqref{C1y} in the large-$N$ limit, taking $s_z=-N/2\cos(\theta(t))$, is
\begin{equation}
\begin{aligned}
C_{1y}(t)
={}&
\frac{\sqrt{N}}
{4\cos^2\left(\theta_0/2\right)}
\Bigg\{
-\operatorname{cn}\left(
u(t),\sin\frac{\theta_0}{2}
\right)
\operatorname{sn}\left(
u(t),\sin\frac{\theta_0}{2}
\right)
\operatorname{dn}\left(
u(t),\sin\frac{\theta_0}{2}
\right)
\\
&\quad+
\Bigg[
A(t)\operatorname{cn}\left(
u(t),\sin\frac{\theta_0}{2}
\right)
+
\operatorname{sn}\left(
u(t),\sin\frac{\theta_0}{2}
\right)
\operatorname{dn}\left(
u(t),\sin\frac{\theta_0}{2}
\right)
\Bigg]
\\
&\qquad\times
\Bigg[
\left(
\cos^2\frac{\theta_0}{2}
-
\sin^2\frac{\theta_0}{2}
\operatorname{sn}^2\left(
u(t),\sin\frac{\theta_0}{2}
\right)
\right)
\operatorname{cn}\left(
u(t),\sin\frac{\theta_0}{2}
\right)
\\
&\hspace{47mm}
-
A(t)\operatorname{sn}\left(
u(t),\sin\frac{\theta_0}{2}
\right)
\operatorname{dn}\left(
u(t),\sin\frac{\theta_0}{2}
\right)
\Bigg]
\Bigg\},
\\[3mm]
C_{11}(t)
={}&
\frac{1}
{4\cos^2\left(\theta_0/2\right)}
\Bigg\{
\operatorname{cn}^2\left(
u(t),\sin\frac{\theta_0}{2}
\right)
\\
&\quad+
\Bigg[
A(t)\operatorname{cn}\left(
u(t),\sin\frac{\theta_0}{2}
\right)
+
\operatorname{sn}\left(
u(t),\sin\frac{\theta_0}{2}
\right)
\operatorname{dn}\left(
u(t),\sin\frac{\theta_0}{2}
\right)
\Bigg]^2
\Bigg\},
\\[3mm]
C_{yy}(t)
={}&
\frac{N}
{4\cos^2\left(\theta_0/2\right)}
\Bigg\{
\operatorname{sn}^2\left(
u(t),\sin\frac{\theta_0}{2}
\right)
\operatorname{dn}^2\left(
u(t),\sin\frac{\theta_0}{2}
\right)
\\
&\quad+
\Bigg[
\left(
\cos^2\frac{\theta_0}{2}
-
\sin^2\frac{\theta_0}{2}
\operatorname{sn}^2\left(
u(t),\sin\frac{\theta_0}{2}
\right)
\right)
\operatorname{cn}\left(
u(t),\sin\frac{\theta_0}{2}
\right)
\\
&\hspace{47mm}
-
A(t)\operatorname{sn}\left(
u(t),\sin\frac{\theta_0}{2}
\right)
\operatorname{dn}\left(
u(t),\sin\frac{\theta_0}{2}
\right)
\Bigg]^2
\Bigg\}.
\label{anal}
\end{aligned}
\end{equation}
Here, $u(t)$ and $A(t)$ are defined as
\begin{equation}
\begin{aligned}
u(t)
&=
K\left(\sin\frac{\theta_0}{2}\right)
-g\sqrt{N}\,t,
\\
A(t)
&=
-g\sqrt{N}\,t
\cos^2\frac{\theta_0}{2}
+
E\left(\sin\frac{\theta_0}{2}\right)
-
\mathcal{E}\left(
u(t),\sin\frac{\theta_0}{2}
\right),
\end{aligned}
\end{equation}here $K,E$ are complete elliptic integrals of the first and second kinds, respectively, and $\mathcal{E}$ is the Jacobi epsilon function. The functions sn, cn, and dn are the Jacobi elliptic sine, cosine, and delta-amplitude functions, respectively. The spin-squeezing parameter is $\xi^2=\frac{4}{N}\min({C_{yy},NC_{11}})$, and the light-squeezing parameter is $r=-\frac{1}{2}\ln\left(4\min(C_{11},C_{yy}/N)\right)$.

\begin{figure*}[t]
\centering
\includegraphics[width=0.8\textwidth]{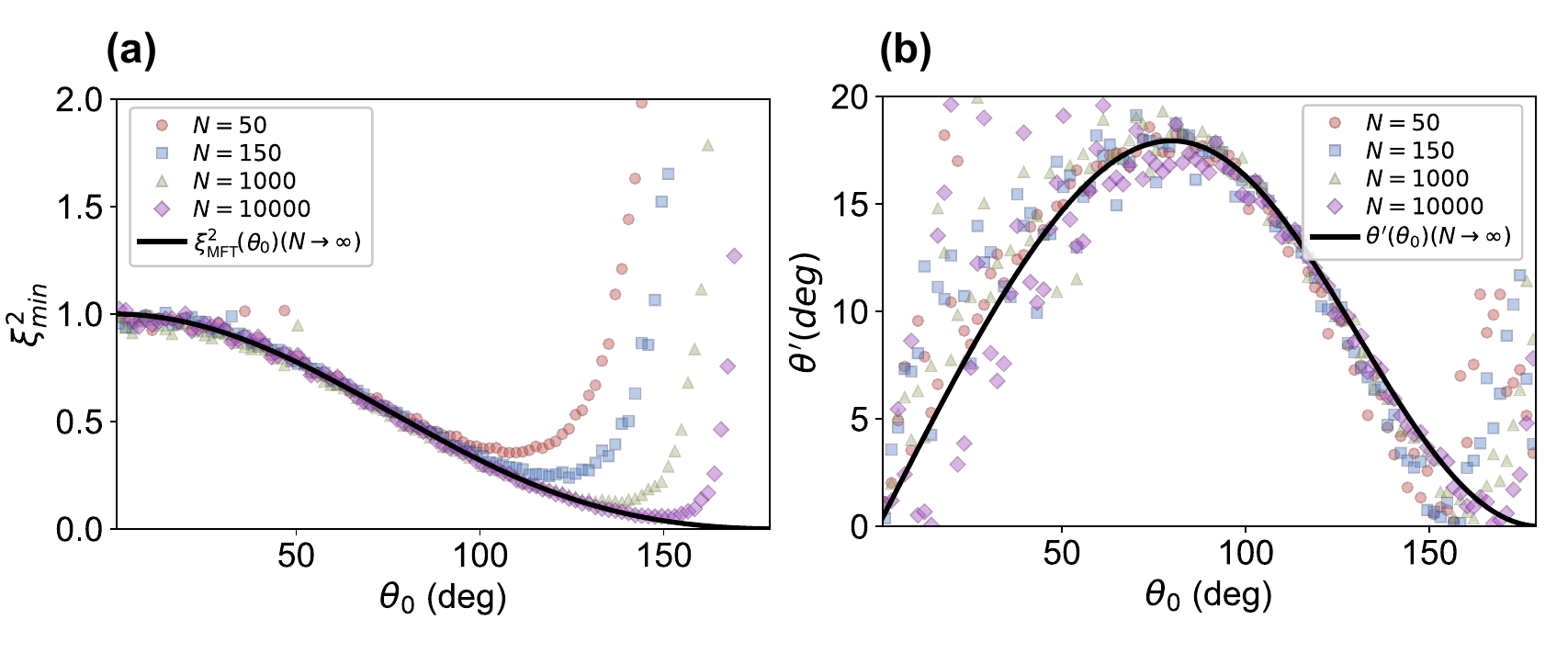}
\caption{\textbf{Comparison of the first-squeezing parameter $\xi^2_{min}$ and the squeezing position $\theta'$ obtained from DTWA and second-order mean-field theory in the large-$N$ limit.}
(a) Dependence of the squeezing parameter at the first squeezing point on the initial angle $\theta_0$. DTWA results are shown for $N=50,150,1000,10000$. As $N$ increases, the exact DTWA solution approaches the large-$N$ second-order mean-field result $\xi^2_{MFT}(\theta_0)$.
(b) Comparison of the optimal squeezing position predicted by large-$N$ second-order mean-field theory with the DTWA results corresponding to panel~(a). As $N$ increases, the numerical curves approach $\theta'(\theta_0)$.}
\label{supp6}
\end{figure*}

From Eq.~\eqref{anal}, one can derive the squeezing parameter at the first squeezing point, $\xi^2_{MFT}(\theta_0)$, and the corresponding squeezing position, $\theta'(\theta_0)$, as shown in Fig.~\ref{supp6}. For an initial angle $\theta_0$ smaller than the optimal initial angle for a given $N$, second-order mean-field theory agrees well with the DTWA results. It fails when $\theta_0$ is too close to $180^{\circ}$, where the spin--light entanglement becomes very strong and the second-order theory breaks down. As $N\xrightarrow{}\infty$, the higher-order cumulants vanish and the exact solution converges to the second-order solution. Consequently, the curves of $\xi^2_{min}$ and $\theta'$ approach $\xi^2_{MFT}(\theta_0)$ and $\theta'(\theta_0)$.

Figure~\ref{supp6} illustrates the scaling of $\xi^2_{min}$, the corresponding optimal initial angle $\theta_0$, and the squeezing position $\theta'$. The finite-$N$ contribution $\xi^2(\theta_0,N)$ in the absence of vacuum fluctuations approaches zero as $N\xrightarrow{}\infty$. The total squeezing parameter can therefore be estimated as $\xi^2_{min}=\xi^2_{MFT}(\theta_0)+\xi^2(\theta_0,N)$. As $N\xrightarrow{}\infty$, the optimal initial angle approaches $\pi$, while the corresponding squeezing position approaches zero, as shown in Fig.~\ref{supp6}(b).

One can also derive the scaling law of the squeezing parameter. To extract the asymptotic scaling, we consider the limit in which the optimal initial angle approaches $\pi$. We define $\delta=\pi-\theta_0, c=\cos\frac{\theta_0}{2}\simeq\frac{\delta}{2},
\delta\ll1.$ In this limit, the first positive zero of $C_{1y}$ occurs near $u=0$. Expanding the Jacobi elliptic functions in Eq.~\eqref{anal}, the condition $C_{1y}(t_0)=0$ gives $u(t_0)\simeq\frac{c^2}{2}.$ Substituting this result into $C_{yy}(t_0)$, the squeezing contribution from mean-field theory in the large-$N$ limit becomes
\begin{equation}
\xi^2_{MFT}(\theta_0)
=
\frac{4C_{yy}(t_0)}{N}
\simeq
\frac{c^2}{2}
\simeq
\frac{\delta^2}{8}.
\label{asymptotic_intrinsic}
\end{equation}

On the other hand, the finite-$N$ contribution in the absence of vacuum fluctuations in Eq.~\eqref{eq:squeezing_no_vacuum} is dominated near $\theta_0=\pi$. Using
$E(\sin(\theta_0/2))\rightarrow1$,
$\cos(\theta_0/2)K(\sin(\theta_0/2))\rightarrow0$, and
$\sin\theta_0\simeq\delta$, one obtains
\begin{equation}
\xi^2(\theta_0,N)
\simeq
\frac{4}{N\delta^4}.
\label{asymptotic_vacuum}
\end{equation}
Therefore, the total squeezing parameter can be approximated by $\xi^2_{min}
\simeq
\frac{\delta^2}{8}
+
\frac{4}{N\delta^4}.$ The first term decreases as $\theta_0$ approaches $\pi$, whereas the second term increases because of finite-$N$ fluctuations. Minimizing $\xi^2_{min}$ with respect to $\delta$ gives
\begin{equation}
\delta_{\mathrm{opt}}
=
2N^{-1/6},
\qquad
\theta_{0,\mathrm{opt}}
\simeq
\pi-2N^{-1/6},
\end{equation}
which means that the optimal initial angle approaches $\pi$ with the scaling law $N^{-1/6}$. Substituting $\delta_{\mathrm{opt}}$ back into $\xi^2_{min}$, we obtain
\begin{equation}
\xi^2_{opt}
\simeq
\frac{3}{4}N^{-1/3}.
\label{squeezing_scaling}
\end{equation}

Thus, the minimum squeezing parameter follows the asymptotic power law $\xi^2_{opt}\propto N^{-1/3}$. Moreover, since $\theta'(\theta_0)\simeq2u(t_0)\simeq c^2$, the optimal squeezing position scales as
\begin{equation}
\theta'_{\mathrm{opt}}
\propto N^{-1/3}.
\end{equation}
These results show that increasing $N$ simultaneously improves the squeezing, shifts the optimal initial angle toward $\pi$, and moves the first squeezing point toward $\theta'=0$.

\section{Exact unitary simulation and calculation of QFI}

In the absence of dissipation, an $N$-qubit system homogeneously coupled to the cavity without external driving is described solely by the Tavis--Cummings interaction term in the resonant rotating frame:
\begin{equation}
    \hat{H}_{\mathrm{TC}} = \hbar g \left( a \hat{S}_+ + a^\dagger \hat{S}_- \right)
    \label{TCH}
\end{equation}

The $U(1)$ symmetry implies conservation of the total number of spin excitations plus photons:
\begin{equation}
    \hat{N}_e = \hat{S}_z + a^\dagger a
\end{equation}

Therefore, for an initial product state consisting of an $n$-photon Fock state and a Dicke state, \(|n\rangle |S, m\rangle\), the evolved state is confined to the subspace satisfying \(N_e = n + m\). It can be expanded in terms of all states satisfying \(n' +m' = n + m\):
\begin{equation}
\exp\left(-ig \left( a \hat{S}_+ + a^\dagger \hat{S}_- \right)t\right) |n\rangle |S, m\rangle
=\sum_{n'}c_{(n', n, m)}(t) \, |n'\rangle |S,  n + m-n'\rangle.
\end{equation}

For the initial state $|CSS\rangle\otimes|0\rangle$, the general evolved state can be written as a superposition in the Dicke subspace:
\begin{equation}
    |\psi(t)\rangle= \sum_{n'=0}^{N}\sum_{m'=-S}^{S}F_{(n', m')}(t) \, |n'\rangle |S, m'\rangle,
\end{equation}

The photon number is truncated because of energy conservation. The coefficients \(F_{(n,m)}(t)\) are determined by the Schr\"odinger equation generated by the TC Hamiltonian in Eq.~\eqref{TCH}, and the Hilbert space separates into several independently evolving chains:
\begin{equation}
i \dot{F}_{(n',m')}
=
g \sqrt{n'+1} \sqrt{(S+m')(S-m'+1)} \, F_{(n'+1, m'-1)}
+
g \sqrt{n'} \sqrt{(S-m')(S+m'+1)} \, F_{(n'-1, m'+1)},
\end{equation}
with $n' = 0,1,2,\dots, N$, $m' = -S, -S+1, \dots, S$, and the initial condition
$F_{(n,m)}(0)
= \delta_{n0} \binom{N}{S-m}^{1/2}
\left( \cos\frac{\theta_0}{2} \right)^{S-m}
\left( \sin\frac{\theta_0}{2} \right)^{S+m}$.

If only the spin information is required, the reduced spin density matrix is obtained by tracing out the light mode:
\begin{equation}
    \hat{\rho}_s=\sum_{m,m'}\sum_{n}\left(F_{(n,m)}(t)F^*_{(n,m')}(t)\right)|m\rangle\langle m'|.
\end{equation}

The quantum Fisher information (QFI) is an important concept in quantum metrology. It is related to the Cram\'er--Rao bound $\Delta\theta\ge\Delta \theta_{QCR}=\frac{1}{\sqrt{\nu F_Q[\hat{\rho}]}}$ for $\nu$ independent measurements of the phase $\theta$ and represents the optimal phase sensitivity of a given probe state and interferometric transformation~\cite{pezze_quantum_2018}. For an $N$-qubit system, if the density matrix is diagonalized as $\hat{\rho}=\sum_{k}q_k|k\rangle \langle k|$, the QFI $F_Q[\hat{\rho},J_n]$ is defined as the largest eigenvalue of the matrix $[\Gamma_Q]_{ij}$~\cite{hyllus_not_2010}:
\begin{equation}
    \left[\mathbf{\Gamma}_Q\right]_{ij} = 2 \sum_{q_k+q_{k'} > 0} \frac{(q_k - q_{k'})^2}{q_k + q_{k'}} \langle k' | \hat{J}_i | k \rangle \langle k | \hat{J}_j | k' \rangle.
\end{equation}

For a pure state, $[\Gamma_Q]_{ij}$ reduces to the covariance matrix of the spin operators, $[\Gamma_Q]_{ij}=4Cov[\hat{\rho},J_n]_{ij}$. Hence, the QFI is given by the maximal spin variance of the spin state. For separable $N$-qubit states, $F_Q[\hat{\rho},J_n]\le N$~\cite{pezze_entanglement_2009}, and the optimal phase sensitivity achievable with a CSS is $\Delta
\theta_{SQL}=1/\sqrt{\nu N}$, which is the standard quantum limit~\cite{giovannetti_quantum_2006}.

The condition $F_Q[\hat{\rho},J_n]>N$ is sufficient for particle entanglement useful in quantum metrology~\cite{pezze_entanglement_2009}. The relation between the QFI and the squeezing parameter follows from the inequality $\Delta\theta\ge\Delta\theta_{QCR}$ and the definition of the Wineland parameter $\xi^2=\frac{N(\Delta J_n)^2}{|J|^2}$, which give $F_Q[\hat{\rho},J_n]/N\ge1/\xi^2$~\cite{pezze_entanglement_2009}. This inequality can also be interpreted as a bound on the metrological enhancement. Therefore, $\xi^2<1$ is a sufficient condition for useful entanglement.

\clearpage

\end{document}